\documentclass[pdftex,twocolumn,epjc3]{svjour3}          
\RequirePackage[T1]{fontenc}
\smartqed  
\RequirePackage{graphicx}
\RequirePackage{mathptmx}      
\RequirePackage{flushend}
\RequirePackage[numbers,sort&compress]{natbib}
\RequirePackage[colorlinks,citecolor=blue,urlcolor=blue,linkcolor=blue]{hyperref}
\journalname{Eur. Phys. J. C}

\usepackage{bbold}
\usepackage[cmtip,arrow]{xy}
\usepackage{pb-diagram,pb-xy}
\usepackage{slashed}
\usepackage{graphicx}
\usepackage{amssymb}
\usepackage{float}
\usepackage{amsopn}
\usepackage{fancyhdr}
\usepackage[yyyymmdd,hhmmss]{datetime}
\newcommand{\be}{\begin{equation}}
\newcommand{\ee}{\end{equation}}
\newcommand{\en}{\end{equation}}
\newcommand{\ba}{\begin{eqnarray}}
\newcommand{\ea}{\end{eqnarray}}
\newcommand{\bea}{\begin{eqnarray}}
\newcommand{\eea}{\end{eqnarray}}
\newcommand{\pa}{\partial}

\def\ps{p\!\!\!/}

\def\ns{n\!\!\!/}

\def\ds{\partial\!\!\!/}

\def \be {\begin{equation}}
\def \ee {\end{equation}}
\def \bea {\begin{eqnarray}}
\def \eea {\end{eqnarray}}
\def \sla {\slashed}
\begin{document}
\title{Renormalization in a Lorentz-violating model and higher-order operators }

\author{J. R. Nascimento\thanksref{addr1,e1}
        \and A. Yu. Petrov\thanksref{addr1,e2} \and Carlos M. Reyes\thanksref{addr2,e3} }
\thankstext{e1}{e-mail: jroberto@fisica.ufpb.br}
\thankstext{e2}{e-mail: petrov@fisica.ufpb.br}
\thankstext{e3}{e-mail: creyes@ubiobio.cl}

\institute{Departamento de F\'{\i}sica, Universidade Federal da Para\'{\i}ba
 Caixa Postal 5008, 58051-970, Jo\~ao Pessoa, Para\'{\i}ba, Brazil  \label{addr1}
          \and
          Departamento de Ciencias B\'{a}sicas, Universidad del B\'{\i}o B\'{\i}o,
Casilla 447, Chill\'{a}n, Chile\label{addr2}
       }


\date{Received: date / Accepted: date}
\maketitle
\begin{abstract}
The renormalization in a Lorentz-breaking scalar-spinor 
higher-derivative model involving $\phi^4$ 
self-interaction and the Yukawa-like coupling is studied. 
We explicitly de- monstrate that the convergence 
is improved in comparison with the usual scalar-spinor model, so, 
the theory is super- renormalizable, with no divergences 
beyond four loops. We compute the one-loop corrections to the
propagators for the scalar and fermionic fields
 and show that in the presence of higher-order
Lorentz invariance violation, the poles that dominate the physical theory, are 
driven away from the standard on-shell pole mass due to radiatively induced lower dimensional operators.
The new operators change the standard gamma-matrix structure of the two-point functions, introduce large 
Lorentz-breaking corrections and
lead to modifications in the renormalization conditions of the theory. We found the physical pole mass
 in each sector of our model.
\end{abstract}
\section{Introduction}
It is well known that the Lorentz-breaking field theory models 
can be introduced in several ways. We can 
list some of the most popular approaches. First, one can introduce 
small Lorentz-breaking modifications of the known theories 
through additive terms, thus implementing the Lorentz-breaking extensions of the 
standard model~\cite{ColKost}. In principle, the most known extensions of the QED 
follow this way. A very extensive list of the possible Lorentz-breaking 
additive terms in different field theory models including QED is given 
by~\cite{KosGra}. Second, one can start with the modified dispersion relations~\cite{Amelino}, 
and, in principle, try to find a theory yielding such relations. Third, the Lorentz-breaking 
theories can be treated as a low-energy limit of some fundamental theories, for example 
string theory~\cite{KostSam} and loop quantum 
gravity~\cite{LQG}. Finally, the Lorentz symmetry can be broken 
spontaneously, see f.e.~\cite{spont}. The main motivation behind all 
these approaches, however, is 
the same, and resides in the expectation that any experimental evidence of departure from
Lorentz symmetry may provide the first germs towards the construction of
a theory amalgamating both 
General Relativity and the Standard Model of particle physics.

At the same time, it is natural to consider one more aspect of studying the Lorentz-breaking 
extensions of the field theory models. It consists in introducing essentially Lorentz- breaking
terms, that is, those ones proportional to some constant vectors or tensors, involving higher 
derivatives. As a result, the corresponding theory will yield an essentially different quantum 
dynamics. The first known example of such a theory is the Myers-Pospelov extension of the 
electrodynamics~\cite{MP} where the three-derivative term essentially involves the 
Lorentz symmetry breaking. Another important example of such a theory is the
four-dimensional Chern-Simons modified gravity with the Chern-Simons coefficient chosen 
in special form $\Theta(x)=k_{\mu}x^{\mu}$~\cite{JaPi}, which, in the weak field limit, 
also involves third order in derivatives of the 
dynamical field (that is, the metric fluctuation). 
Moreover, the importance of the Myers-Pospelov-like term, 
and analogous terms for scalar and spinor fields which can be
easily introduced, is also motivated by the fact that a special choice of the Lorentz-breaking 
vector will allow to eliminate the presence of higher time derivatives thus avoiding the 
arising of the ghosts which are typically present in theories with 
higher time derivatives (see f.e.~\cite{ghosts}).
Also, this term was shown to 
arise as a quantum correction in different Lorentz-breaking extensions of QED~\cite{MNP}
and has been studied for causality and stability~\cite{CMR0}.
In the case of including higher time derivatives, it has been shown 
recently that the unitarity of the $S$-matrix can be preserved at the one-loop 
order in a Myers-Pospelov QED~\cite{CMR}.
The proof has been accomplished using the Lee-Wick
prescription for quantum field theories with negative 
metric~\cite{Lee-Wick}. For other studies on unitarity at tree level for minimal
and nonminimal Lorentz violations, see~\cite{Schreck1,Schreck2} respectively. 
 It is important to notice that the Myers-Pospelov-like 
modifications of QED 
are actually experimentally studied as well within different contexts~\cite{MPexp}.

We emphasize that, up to now, the quantum impact of the 
Myers-Pospelov-like class of terms being introduced already 
at the classical level, where the higher-derivative 
additive term should carry a small parameter 
which can enforce large quantum corrections~\cite{collins}, almost 
was not studied except of the QED~\cite{Fine-Tuning} and superfield 
case~\cite{CMP}. The presence of such effect raises the question 
 how to define correctly the physical parameters in the renormalized theory. 
On the other hand, for studies in the context of semiclassical quantization
it is natural to consider   
the presence of higher derivative terms in order
to implement a consistent renormalization program~\cite{shapiro}.
With these considerations, the natural question is -- what are the possible consequences of 
including the Lorentz-breaking higher-derivative terms into the classical action?

It is well known that loop corrections in Lorentz- violating
quantum field theory may lead to new kinetic operators absent in the 
original Lagrangian. Recently,
the consequences of these radiatively induced operators have been studied in relation with the finiteness 
of the $S$-matrix and the identification of the asymptotic state space~\cite{ralf}.
These new terms introduce modifications in the propagation of free particles
and change drastically the physical content of the space of in and out states.
In particular, the K\"all\'en--Lehmann representation~\cite{KL} 
and the Lehmann- Symanzik- Zimmermann (LSZ) reduction 
formalism~\cite{LSZ} are modified in the presence 
of Lorentz symmetry violation~\cite{rob}.
An important finding is that spectral densities which in the standard case are functions of 
momentum-dependent 
observer scalars such as $p^2$, in the Lorentz violating scenario may depend on other scalars such as 
couplings of Lorentz-violating tensor coefficients with momenta~\cite{rob}.
This has led to modifications in the renormalization procedure,
in the definition of the asymptotic Hilbert space and in general in the
treatment for external-leg physics~\cite{ralf}; for other studies of the
renormalization in Lorentz-breaking theories, see also~\cite{scarp}.  
A natural extension for these studies is to consider  
the nonminimal framework of Lorentz invariance violation, that is, when the Lorentz-breaking 
is performed with higher-order operators~\cite{Kos-Mew}.
It is well known that the inclusion of 
higher-order operators in quantum field theory will generate, via radiative corrections,
all the lower dimensional operators allowed by the symmetries of the Lagrangian~\cite{gen_rad}.
For the case of breaking the Lorentz symmetry, let us say in QED and with a 
preferred four-vector $n^{\mu}$, 
the induced operators may involve
contractions of $n_{\mu}$ with matrices other than just $\gamma^{\mu}$,
together with scalars such as $(n \cdot p)$. 
The new terms force to modify
 the renormalization conditions in order to extract the correct pole mass from the two-point functions.
In particular, the renormalization condition for the renormalized fermion self-energy
$\Sigma_R(\sla{p}=m_{P})=0$, with $m_{P}$ being the physical pole 
mass, has to be generalized, which ultimately
will depend on the form of  Lorentz breaking.
In this work we continue these studies in order to carry out
the renormalization in a theory with higher-order operators and in addition we study the
 possible effects of large Lorentz-violating corrections.
Within our study, we consider the renormalization of the higher-derivative 
Lorentz-breaking generalizations of $\lambda \phi^4$ and Yukawa model. 

The structure of the paper looks like follows. In Sec.~\ref{Sec2},
we consider the classical actions of our 
models, write down the dispersion relations, find the poles and describe their analytical 
behavior in complex $p_0$-plane.
In Sec.~\ref{Sec3}, we compute the quantum corrections corresponding to the self 
interaction $\lambda \phi^4$. In Sec.~\ref{Sec4} we discuss the coupling of
scalar and spinor fields and provide a study of the degree of
divergencies in our model. In Sec.~\ref{Sec5}, we compute the two-point 
functions in purely scalar and scalar-spinor sectors
 thus exhausting possible divergences and showing explicitly the radiatively 
 induced operators with new gamma-matrix structure and large Lorentz-violating terms.
In Sec.~\ref{Sec6}, we perform the mass renormalization in both sectors and find the
physical masses in the theory. In the last section, we discuss 
our results, and in the Appendices~\ref{AppendixA},~\ref{AppendixB} and \ref{AppendixC},
we provide some details of the calculations.
\section{The effective models and pole structure}\label{Sec2}
We are interested in the higher-order Lagrangian density
describing two sectors of Lorentz-breaking theory
\begin{eqnarray}\label{L_1-L_2}
\mathcal L= \mathcal L_{1}+\mathcal L_2\,.
\end{eqnarray}
The first sector involves a scalar sector with a fourth derivative
together with a self-interaction potential term
\begin{eqnarray}
\mathcal L_{1}=\frac{1}{2}\partial_{\mu} \phi \partial^{\mu}\phi  
-\frac 1 2M^2\phi^2 +g_1\phi (n\cdot \partial)^4\phi -\frac{\lambda}{ 4!}\phi^4\,,
\end{eqnarray}
and the second one the fermionic Myers-Pospelov model~\cite{MP},
with dimension-five operators and the Yukawa coupling vertex
\begin{eqnarray}\label{yukawa}
{\cal L}_2=\bar{\psi}\left(i\ds-m-\bar{\alpha} m\ns  \right)\psi +g_2\bar{\psi}\ns(n\cdot\pa)^2\psi  
+g\bar{\psi} \phi \psi\,.
\end{eqnarray}
The constants $g_1=\frac{\kappa}{M^2_{Pl}}$ and
$g_2=\frac{\eta}{M_{Pl}}$ parametrize the higher-order 
Lorentz invariance violation with $M_P$ 
 the Planck mass representing itself as a natural mass scale, 
$\kappa$, $\eta$ and $\bar{\alpha}$ are dimensionless parameters, whose presence 
describes the intensity of the higher-derivative terms, and $n^{\mu}$ is 
a dimensionless four-vector defining a preferred reference frame.
\begin{figure}
\centering
\includegraphics[width=0.5 \textwidth]{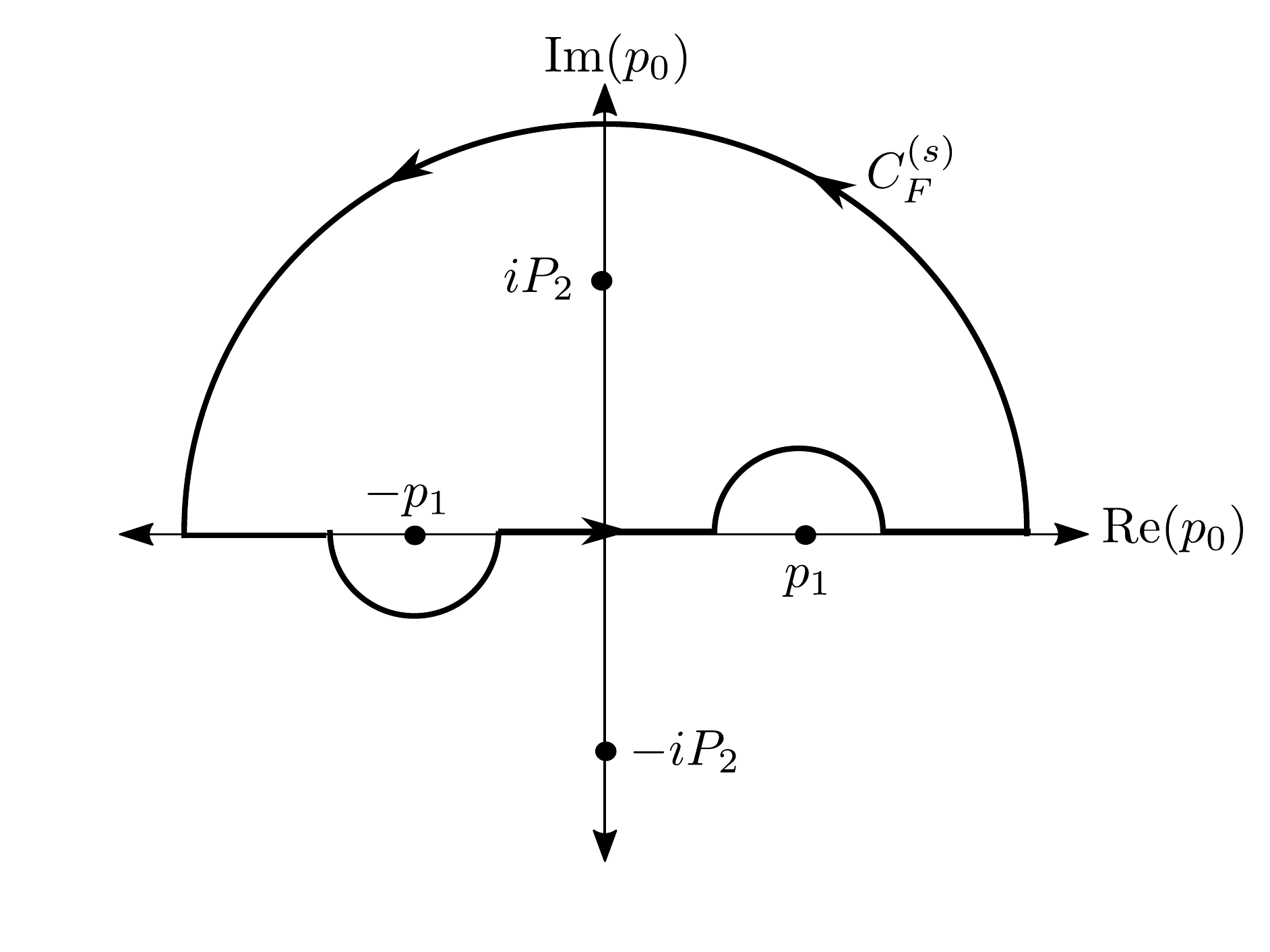}
\caption{\label{Fig1}  
The path of integration $C_F^{(s)}$ for the scalar propagator which 
encloses the poles $- p_1$ and $iP_2$ when closing the contour upward and encloses the 
poles $p_1$ and $-iP_2$ when closing the contour downward. 
} 
\end{figure}

The propagators in momentum space read
\begin{eqnarray}\label{propagators}
\Delta(p)&=&    \frac{i}{p^2-M^2+2g_1 (n\cdot p)^4}\,,\nonumber
\\
S(p)&=&\frac{i}{\ps-m-\bar{\alpha} m\ns-g_2\ns (n\cdot p)^2   }\,.
\eea
We begin an analysis of the dispersion relations in both sectors.
A further 
motivation for its study, and consequently, the 
finding of the poles and their analytical behavior in complex $p_0$-plane,
consists first in the fact that in our models
 namely using of the residues of the propagators is a most 
 convenient approach for calculating the quantum corrections. Second,
in the presence of higher-order time-derivative terms a direct implementation
of the $i\epsilon$ prescription may lead to a wrong four-momentum
representation for the propagator which may spoil any attempt to preserve 
unitarity or causality.

Let us start with the scalar dispersion relation, namely
\begin{eqnarray}
p^2-M^2+2g_1(n\cdot p)^4=0\,,
\end{eqnarray}
which for a purely time-like four-vector $n^{\mu}=(1,0,0,0)$, the 
solutions are given by
\begin{eqnarray}
p_0=\pm \frac{1}{2}\sqrt{\frac{-1\pm \sqrt{1+8g_1E^2}}{g_1}},
\end{eqnarray}
and where $E(\vec p)=\sqrt{{\vec p}^2+M^2}$. 
The dispersion relation can also be written as $(p_0^2-p_1^2 ) (p_0^2+P_2^2 )=0$,
hence one has the solutions $p_0=\pm p_1$ and $p_0=\pm i P_2$ so that
\begin{eqnarray} \label{scalar-sol}
p_{1}&=&\frac{1}{2}\sqrt{\frac{-1+ \sqrt{1+8g_1E^2}}{g_1}}\,, \nonumber  \\
P_{2}&=&\frac{1}{2}\sqrt{\frac{1+ \sqrt{1+8g_1E^2}}{g_1}}\,.
\end{eqnarray}
Their exact location in complex $p_0$-plane and also
 the contour of integration $C_F^{(s)}$
are shown in Fig.~\ref{Fig1}.

The solutions can be classified according to their perturbative behavior when 
taking the Lorentz violation to zero.
We identify two standard solutions $\pm p_1$ which are
perturbative solutions to the usual ones $\pm E$ and 
two complex ones (and moreover, actually tachyonic) $\pm i P_2$ 
 which diverge as $g_1 \to 0$. The extra solutions that appear
 $\pm P_2$ are associated to negative-metric states in Hilbert space
 and have been called Lee-Wick solutions~\cite{Lee-Wick}.
\begin{figure}
\centering
\includegraphics[width=0.5 \textwidth]{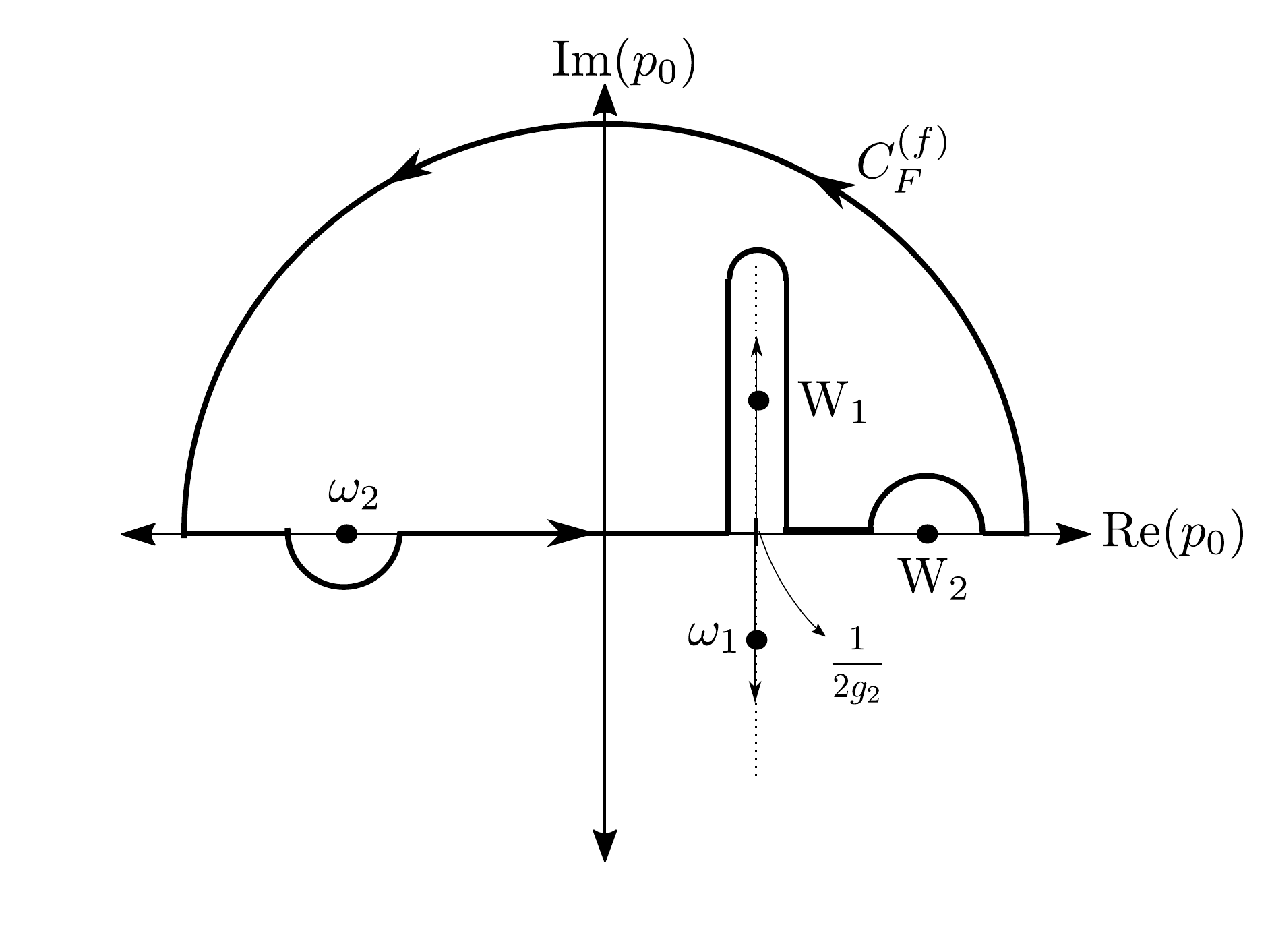}
\caption{\label{Fig2}  
The integration contour $C_F^{(f)}$ for the fermion propagator which 
is defined to round the negative 
pole from below and the positive poles from above.
At higher energies than $1/(4g_2)$,
the two solutions $\omega_1$ and $W_1$ become complex and move in opposite 
directions along the imaginary line starting at $1/(2g_2)$, 
we deform the contour
 continuously avoiding any crossing or singularity with the poles.} 
\end{figure}

Alternatively, we can write the scalar propagator as
\begin{eqnarray}
\Delta(p)&=&    \frac{i}{2g_1 (p_0^2+P_2^2)(p_0^2-p_1^2+i\epsilon) }\,,
\eea
which agrees with the usual propagator in the limit $g_1\to 0$. 

In the fermion sector we have the dispersion relation 
\begin{eqnarray}
(p_ {\mu}-\bar{\alpha} m n_{\mu}-g_2n_ {\mu} (n\cdot p)^2  )^2  -m^2=0\,.
\end{eqnarray}
Again for the time-like $n^{\mu}$ we have the equation
\begin{eqnarray}
(p_ 0-\bar{\alpha}m -g_2p_0^2  )^2  -\vec p^2-m^2=0\,,
\end{eqnarray}
whose standard, that is, non-singular at $g_2\to 0$, solutions are 
\begin{eqnarray} \label{field-energy}
\omega_1&=&\frac{ 1-\sqrt{1-4g_2(\bar{\alpha} m+\mathcal{E})}}{2g_2}\nonumber\\
\omega_2&=&\frac{ 1-\sqrt{1+4g_2 (\mathcal{E}-\bar{\alpha} m)}}{2g_2}\,, 
\end{eqnarray}
and the Lee-Wick ones
\begin{eqnarray}
W_1&=&\frac{ 1+\sqrt{1-4g_2(\bar{\alpha} m+\mathcal{E}) }}{2g_2}\nonumber\\
W_2&=&\frac{ 1+\sqrt{1+4g_2 (\mathcal{E}-\bar{\alpha} m) }}{2g_2}\,,
\end{eqnarray}
where $\mathcal{E}(\vec p)=\sqrt{\vec p^2+m^2}$. 

In the region of energies satisfying the condition $4g_2(\mathcal{E}\pm\bar{\alpha})<1$ 
the four solutions are real and obey the inequality
 $\omega_2<\omega_1< W_1<W_2$ at least for $\bar{\alpha}$ enough small, where $\omega_2$ is a negative number.
However, beyond the critical energy $1/(4g_2)$ both
$\omega_1$ and $W_1$ become complex and move in the opposite 
imaginary line at $1/(2g_2)$ as shown in 
Fig.~\ref{Fig2}, while the other two solutions $\omega_2, W_2$ remain real.

To define the contour $C_F^{(f)}$ we use an heuristic argument, specially
to go beyond the critical energy at which complex solutions appear. We implement
a correct low energy limit by considering the prescription given 
in~\cite{QEDunitarity} which has been well tested to give a suitable
correspondence with the normal theory
when  $g_2\to 0$ and also to preserve the unitarity of the $S$ matrix.
In this effective region the integration contour $C_F^{(f)}$ is defined 
to round the negative pole from below 
and the three positive ones from above. 
Now we increase the energy to values at which 
 the two solutions $\omega_1$ and $W_1$ become 
complex, and define the new contour as the one obtained by continuously deforming 
 the curve by avoiding any crossing and singularity with the 
 poles, as shown in Fig.~\ref{Fig2}.

With this consideration in mind, the fermion propagator reads
\begin{eqnarray}
S(p)&=&\frac{i\left((p_0-\bar \alpha m-g_2 p_0^2)\gamma^0 +p_i \gamma^i  +m\right)}{g_2^2(p_0-\omega_1+i\epsilon) 
 (p_0-W_1+i\epsilon) (p_0-\omega_2-i\epsilon)}  \nonumber   \\   &\times& \frac{  1}{(p_0-W_2+i\epsilon) }   \,,
\eea
which differs from the direct $i\epsilon$ prescription in the quadratic terms, but allows in particular to
define a consistent Wick rotation
which we use later.
\section{The interaction $\lambda \phi^4$}\label{Sec3}
In this section we explore the potentially divergent 
one-loop radiative correction in the scalar propagator which
is generated 
by the well-known tadpole graph given by 
Fig.~\ref{Fig3}.
\begin{figure}[H]
\centering
\includegraphics[width=0.25 \textwidth]{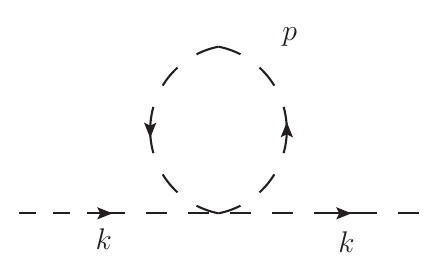}
\caption{\label{Fig3} One-loop graph in the scalar model with self interaction $\lambda \phi^4$.}
\end{figure}

To proceed with it, we need to evaluate the basic integral
\begin{eqnarray}\label{Sigma2}
\Sigma_2=\frac 1 2 \lambda \int \frac{d^4 p}{(2\pi)^4} \frac{1}{p^2-M^2
+2g_1 (n\cdot p)^4 }\,.
\end{eqnarray}
with some regularization technique.

Before computing the above integral, let us comment on the well known fact
that perturbative approximation for physical observables can show some ambiguities 
depending on the regularization scheme. This brief detour will give us some insight of the 
finite nature of some integrals which we will compute using the scheme of 
 analytic continuation to $d$ dimensions or dimensional regularization (DR).

To be more clear, let us consider the integral $\Sigma_2$ in the standard case (taking $g_1\to 0$),
 for which we will arrive to the well known expression in dimensional regularization
\begin{eqnarray}
\Sigma_2^{\rm DR}=\frac{i\lambda M^2}{16\pi^2 \epsilon}+\rm{fin}.
\end{eqnarray}
On the other hand, using a cutoff regularization in which we introduce an upper limit in momenta proportional to $\Lambda$ will produce
\begin{eqnarray}
\Sigma_2^{\rm cut-off}=\frac{\Lambda^2}{2}-\frac{M^2}{4}\left(2\ln
 \left(\frac{2\Lambda}{M}\right)  -1\right)+\mathcal O(\Lambda^{-2})\,,
\end{eqnarray}
which includes quadratic and logarithmic divergencies.

These two results not only show the ambiguity of results for observables in the perturbative scheme, but also
the important fact that quadratic divergencies are not seen by 
dimensional regularization which only has the property to describe logarithmic divergences.
 In our case something similar happens.
As we will show in the Appendix~\ref{AppendixC}, some of our integrals display divergences proportional to powers of $\Lambda$
with the absence of logarithmic divergences, and hence dimensional regularization will give finite 
results for these integrals. 

We return to the computation of $\Sigma_2$. We use DR and go to $d$ dimensions and
choose the Lorentz-breaking four-vector to 
be timelike $n^{\mu}=(1,0,0,0)$ which yields
\begin{eqnarray}
\Sigma_2&=&\frac 1 2 \mu^{4-d}\lambda \int \frac{d^d p}{(2\pi)^d} \frac{1}{
2g_1(p_0^2-p_1^2+i\epsilon)(p_0^2+P_2^2)} \,,
\end{eqnarray}
where $p_1,P_2$ are given in~(\ref{scalar-sol}).
We perform the integration in the complex $p_0$-plane by closing
the contour $C_F^{(s)}$ upward
and enclosing the two poles $-p_1$
and $iP_2$ as depicted in Fig.~\ref{Fig1}, yielding
\begin{eqnarray}\label{sigma_2}
\Sigma_2=\pi i \mu^{4-d}\lambda
 \int \frac{d^{d-1} {\bf p}}{(2\pi)^d}\left( iF_1  -F_{2} \right) \,,
\end{eqnarray}
where 
\begin{eqnarray}
F_1&=&\frac{\sqrt{g_1}}{\sqrt{1+8 g_1E^2}\sqrt{1+\sqrt{1+8 g_1E^2}}}   \,, \nonumber \\
F_{2}&=&\frac{\sqrt{g_1}}{\sqrt{1+8 g_1E^2}\sqrt{-1+\sqrt{1+8 g_1E^2}}} \,.
\end{eqnarray}
Note that $iF_1-F_{2}$
 has the correct limit at $g_1 \to 0$, recovering the usual result $-\frac{1}{2E}$. 
Now, it is convenient to change variables $z=  \sqrt{1+8g_1 E^2({\bf p})}$
yielding
\begin{eqnarray}
{\bf p} \,d{\bf p}=\frac{zdz }{8g_1}; \qquad  
d^{d-1} {\bf p}= |{\bf p}|^{d-2}\, d{\bf p}\; d\Omega_{d-1}  \,,
\end{eqnarray}
which allows to write the integral~(\ref{sigma_2}) as
 \begin{eqnarray}
\Sigma_2= - \pi \mu^{4-d}\lambda
\frac{  2\pi^{(d-1)/2} \sqrt{g_1} }{(2\pi)^d  \Gamma(\frac{d-1}{2}) (8g_1)^{\frac {d -1}{2}}  }
 \, (I_{1}+I_{2})\,.
 \end{eqnarray}
with
\begin{eqnarray}\label{Iuno}
I_1&=& \int _{z_0}^{\infty}
dz \,  \frac{  (z^2-z_0^2)^{\frac{d-3}{2}}  }{\sqrt{z+1}}\,,
\end{eqnarray}
\begin{eqnarray}\label{Ia2}
I_2&=&i \int _{z_0}^{\infty}
dz \,  \frac{  (z^2-z_0^2)^{\frac{d-3}{2}}  }{\sqrt{z-1}}\,,
\end{eqnarray}
where $z_0=\sqrt{1+8g_1M^2}$, and we have used the definition of solid angle (\ref{d-solidangle}).

Considering both contributions through the relation $\Sigma_2=\Sigma_{(1)}+\Sigma_{(2)}$,
and after some algebra with
 $d=4-\epsilon$ and expanding in $\epsilon$, we find
 at the lowest order
 \begin{eqnarray}
 \Sigma_{(1)}&=& \frac{M^2 \lambda }{12\pi^3}   
\left(-\frac{19}{3}+2\gamma_E+6\ln (2)   \right.  \\    &&  \left. -
\frac{  3\pi   \,\, {}_2F_1R^{(0,0,1,0)}  \left(\frac 1 4,\frac 3 4,2,1\right) }
{8\sqrt{2}} +\right.\nonumber\\ &+&\left.
\left(1-\frac{3}{8}\gamma_E-\frac{1}{8}\ln (512)\right) 
 \ln\left(-\frac{g_1M^2}{8}\right)\right)    \,,\nonumber 
\\
\Sigma_{(2)}&=& \frac{\lambda}{144g_1 \pi^3}(-14+6\gamma_E+3 
\ln(32 g_1M^2))    \nonumber    \\    &&    +  \frac{M^2\lambda}{192\pi^3} \left(    8\left(-17
+6\gamma_E+3\ln\left(32g_1M^2\right)   \right) \right. \\
&&\left. -2\left(3+\ln\left(\frac{g_1M^2}{8}\right)\right) \right.\times\nonumber\\ &\times&\left.
\left(-14+6\gamma_E+3 
\ln(32 g_1M^2)\right) \right)\,. \nonumber 
\end{eqnarray}
Here, ${}_2F_1R^{(0,0,1,0)}\left(\frac 1 4,\frac 3 4,2,1\right)$ 
is a hypergeometric function of the given arguments, its value is $-0.71$. 
Note that there is a fine tuning in this case, that 
is, the expression is singular at $g_1\to 0$. However the correction
to the
 two-point function $\Sigma_2$ is UV finite.
\section{Coupling of scalar and spinor fields}\label{Sec4}
Let us consider the theory involving both the quartic interaction 
vertex $V_1=-\frac{\lambda}{4!}\phi^4$ and the Yukawa  
coupling vertex $V_2=g\bar{\psi}\psi\phi$.  
We note, that, in principle, the second time derivatives in a free 
action of a spinor field are present also in specific Lorentz 
invariant theories, for example, the known ELKO model~\cite{ELKO}. 
However, our theory essentially differs from that model.
To classify the possible divergences, we should calculate the
 superficial degree of divergence $\omega$ of this theory. 
 The naive result for it is
\begin{eqnarray}
\omega=4-4V_1-2V_2-E_{\psi}\,,
\end{eqnarray}
where $E_{\psi}$ is a number of spinor legs. However, this 
manner yields incorrect results because of the strong anisotropy
between time and space components of the momenta (for example, 
in this case one can naively suggest that the two-point function of
the spinor field can yield only the renormalization of the mass of the 
spinor field). So, let us proceed in the manner similar to that one used
for Horava-Lifshitz-like theories (cf.~\cite{Anselmi}). Since $n^{\mu}$ 
is purely time-like, we can write $(n\cdot p)^4=p^4_0$, so, we have from 
~(\ref{propagators})
\bea
\Delta(p)&=&    \frac{i}{p^2_0-\vec{p}^2-M^2+2g_1 p^4_0}\,,  \nonumber\\
S(p)&=&\frac{i}{\ps-m-\bar{\alpha} m \gamma_0-g_2\gamma^0p^2_0}\,.
\eea
Following the methodology developed for the 
Horava- Lifshitz theories (see f.e.~\cite{Anselmi}), we suggest that
the denominators of the propagators are the homogeneous functions 
with respect to higher orders in corresponding momenta, 
and the canonical dimension of the spatial momentum $\vec{p}$ is 1. 
Taking into account only the leading degrees, we 
easily conclude that the canonical dimension of the 
momentum $p_0$ is $1/2$ (we note that this case does not occur 
in usual Horava-Lifshitz-like theories where the canonical dimension of time 
momenta are always more than one, cf.~\cite{Anselmi}). Therefore, 
the spinor propagator has the canonical dimension (and the 
contribution to the superficial degree of divergence) equal to $(-1)$, 
and the scalar one -- equal to $(-2)$ just as in the usual case. 
Nevertheless, the dimension of the integral measure, that is, 
$d^4k=d^3\vec{k}dk_0$ in this case is different from the usual 
one, being equal to $7/2$ rather than $4$.
Hence the superficial degree in our theory is
\bea
\omega=(7/2)L-2P_{\phi}-P_{\psi}\,,
\eea 
where $L$ is a number of loops, and $P_{\phi}$ and $P_{\psi}$
 are the numbers of scalar and spinor propagators respectively. 
 Then, let $V_1$ will be the number of $\phi^4$ vertices, and $V_2$ -- 
 of Yukawa-like vertices. One has the identities for numbers of 
 scalar and spinor fields in an arbitrary Feynman diagram:
\bea
N_{\phi}&=&4V_1+V_2=2P_{\phi}+E_{\phi}\,,
\nonumber\\
N_{\psi}&=&2V_2=2P_{\psi}+E_{\psi}\,,
\eea 
where $E_{\phi}$, $E_{\psi}$ are the numbers of 
external scalar and spinor legs respectively.
We use the topological identity $L+V-P=1$, that is, 
$L+V_1+V_2-P_{\psi}-P_{\phi}=1$. As a result, we 
eliminate numbers of loops and propagators from 
$\omega$ and rest with
\bea
\omega=\frac{7}{2}-\frac{1}{2}V_1-\frac{1}{4}V_2-\frac{3}{4}
E_{\phi}-\frac{5}{4}E_{\psi}\,.
\eea
A straightforward verification shows that the superficially 
divergent diagrams (that is, those ones with $\omega>0)$ 
can be of the following types:

(i): $E_{\psi}=2$, $E_{\phi}=0$, $V_2=2$, $\omega=1/2$. 
This is the one-loop renormalization of the mass and kinetic
 terms for the spinor. 

(ii): $E_{\psi}=2$, $E_{\phi}=0$, $V_2=4$, $\omega=0$. 
This is the two-loop renormalization of the mass and kinetic
 terms for the spinor.

(iii): $E_{\psi}=0$, $E_{\phi}=2$, $V_2=2$, $\omega=3/2$. 
This is the one-loop renormalization of the mass and kinetic terms for the scalar. 

(iv): $E_{\psi}=0$, $E_{\phi}=2$, $V_2=4$, $\omega=1$. 
This is the two-loop renormalization of the mass and kinetic terms for the scalar.

(v) $E_{\psi}=0$, $E_{\phi}=2$, $V_1=1$, $\omega=3/2$. 
This is the one-loop renormalization of the mass term for the 
scalar. Actually, we already showed in the previous section that, 
due to the specific structure of poles of the propagator, this 
contribution is finite.

Actually, in the cases (iii) and (iv) the divergence will be 
not linear but logarithmic, by the reasons of symmetry of
 integrals over momenta. The diagrams with odd numbers of 
 $E_{\phi}$ will vanish due to 
an analogue of the Furry theorem.
So, our theory is super-renormalizable.  
Moreover, we note that since the kinetic term for the 
scalar involves two derivatives acting to the external
 fields, its superficial degree of freedom should be 
 decreased at least by 1, if these derivatives are the time 
 ones, and by two for space derivatives; actually, in 
 one-loop case in a purely scalar sector the kinetic 
 term simply does not arise. Also, in the cases (i) and 
 (ii) one will have the only divergent contribution to the 
 mass of the spinor. So, taking into account the 
 previous section as well, we conclude that at the 
 one-loop order one could have only the renormalization 
 of the masses of the spinor and the scalar arisen from 
 the Yukawa-like coupling.

We note that namely this degree of divergence correctly 
explains why the self-energy of the fermion diverges, as 
we will see further (indeed, the naive calculation yields a 
finite result for it).
To study the renormalization, we can restrict ourselves by 
the lower order, that is, one loop.

So, we rest with only three potentially divergent graphs -- 
with $V_1=1$, that is, the purely scalar tadpole we studied 
above, and with $V_2=2$ and $E_{\psi}=0, E_{\phi}=2$ or
 $E_{\psi}=2, E_{\phi}=0$  we study below.
\section{The Yukawa-like theory}\label{Sec5}
In the next subsections we compute the radiative corrections 
to the scalar and fermion two-point function
in the Yukawa-like theory which arises by considering the self-interaction 
term $V_1\to 0$ and $g_1\to 0$ in~(\ref{L_1-L_2}). The Lagrangian is 
 \begin{eqnarray}\label{Ymodel}
\mathcal L&=&\frac{1}{2}\partial_{\mu} \phi \partial^{\mu}\phi  
-\frac 1 2M^2\phi^2  
+\bar{\psi}\left(i\ds-m -\bar{\alpha} m\ns  \right)\psi \nonumber \\&& +g_2\bar{\psi}
\ns(n\cdot\pa)^2\psi  +g\bar{\psi} \phi \psi\,,
\end{eqnarray}
and additionally, we impose the simplification of considering 
$m=M$ and the preferred four-vector to be purely timelike $n=(1,0,0,0)$.
\subsection{Scalar self-energy  $\Pi(p)$ }\label{subSec5}
As a first example of quantum corrections in our Yukawa-like model, 
we study the contribution with two external scalar legs depicted at
Fig.~\ref{Fig4}. 
\begin{figure}[H]
\centering
\includegraphics[width=0.32 \textwidth]{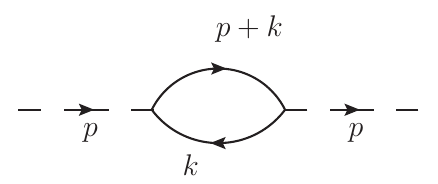}
\caption{\label{Fig4} Self-energy loop graph in the scalar sector.}
\end{figure}
It is represented by the integral
\begin{eqnarray}
i\Pi(p)&=&-\frac{g^2}{2}\phi(-p)\phi(p)\times\nonumber\\ &\times&
\int\frac{d^4k}{(2\pi)^4}
  \frac{ {\rm Tr}  \left((  Q_{\mu} \gamma^{\mu}+m)
(R_{\nu} \gamma^{\nu}+m)\right)}{(Q^2-m^2)(R^2-m^2)}\,,  
\end{eqnarray}
where we define
\begin{eqnarray}\label{QR}
Q_{\mu}&=&k_{\mu}-\bar{\alpha} m n_{\mu}-g_2n_{\mu}(n\cdot k)^2\,, \nonumber\\ 
R_{\mu}&=&k_{\mu}+p_{\mu}-\bar{\alpha} m n_{\mu}-g_2n_{\mu}
(n\cdot (k+p))^2\,.
\end{eqnarray}
Calculating the trace gives
\begin{eqnarray}\label{integral-Pi}
i\Pi(p)=-2g^2\phi(-p)\phi(p)\int\frac{d^4k}{(2\pi)^4}\frac{Q\cdot R+m^2}{
(Q^2-m^2)(R^2-m^ 2)}\,. \nonumber \\
\end{eqnarray}
 In principle, within our calculations, in the denominators $Q^2-m^2$ and $R^2-m^2$ we can suppress the terms
  proportional to $\bar{\alpha}$ since they yield only subleading orders.
Let us write the corresponding contribution to the effective action as $i\Pi(p)=
-2g^2\phi(-p) \widetilde{\Pi}(p)\phi(p)$
and study the typical low-energy behavior 
of this contribution by expanding it into Taylor series
\begin{eqnarray}
\widetilde{\Pi}(p)&=&\widetilde{\Pi}(0)+p_{\mu}\left(\frac{\partial\widetilde{\Pi}}{\partial 
p_{\mu}}\right)_{p=0}+
\frac{1}{2}
p_{\mu}p_{\nu}\left(\frac{\partial^2\widetilde{\Pi}}{\partial p_{\mu}
\partial p_{\nu}}\right)_{p=0}\nonumber  \\ +\dots \,.
\end{eqnarray}
The zeroth-order contribution follows directly from~(\ref{integral-Pi}): 
\begin{eqnarray}\label{first-or}
\widetilde{\Pi}(0)=\int\frac{d^4k}{(2\pi)^4}\frac{Q^2+m^2}{(Q^2-m^2)^2}\,.
\end{eqnarray}
It is convenient to rewrite as 
\begin{eqnarray}
\widetilde{\Pi}(0)=K+2m^2P\,,
\end{eqnarray}
where 
\begin{eqnarray}\label{K}
K=\int \frac{d^4k}{(2\pi)^4}   \frac{1}{(Q^2-m^2)} \,,
\end{eqnarray}
and 
\begin{eqnarray}\label{P}
P=\int \frac{d^4k}{(2\pi)^4}   \frac{1}{(Q^2-m^2)^2}  \,.
\end{eqnarray}
where the integrals (\ref{K}) and (\ref{P}) have been solved in the Appendix~\ref{AppendixA}.

For the next term, it is clear that 
$(\frac{\partial\widetilde{\Pi}}{\partial p^{\mu}})|_{p=0}$
 can be proportional to 
$n_{\mu}$ only since there is no other vectors, 
the corresponding contribution to the effective 
action will yield\\ $\int d^4x\phi(n\cdot\pa)\phi$, 
that is, a surface term. So, we can disregard it.
Further, one would need then to find the second 
derivative, that is, $(\frac{\partial^2\widetilde{\Pi}}{\partial p_{\mu}\partial 
p_{\nu}})|_{p=0}$ which may contain naturally terms of higher-order in $g_2$.
To find it, consider
\begin{eqnarray}
\left(\frac{\partial^2\widetilde{\Pi}(p)}{\partial p_{\mu}\partial p_{\nu}}\right)_{p=0}&=&
\int\frac{d^4k}{(2\pi)^4}{\rm Tr}     \Big[
\left( \frac{1}{\sla{Q}-m}  \right) \nonumber    \\  &\times&  \left[ \frac{\partial}{\partial p_{\mu}} 
  \frac{\partial}{\partial p_{\nu}}  \left(\frac{1}{\sla{R}-m}
\right)\right]_{p=0}
\Big]\,. 
\end{eqnarray}
Integrating by parts and
neglecting the surface terms, we obtain the
symmetric expression
\begin{eqnarray}
\left(\frac{\partial^2\widetilde{\Pi}(p)}{\partial p_{\mu}\partial p_{\nu}}\right)_{p=0}&=&-
\int\frac{d^4k}{(2\pi)^4}{\rm Tr}\Big[  \frac{\partial}  {\partial k_{\nu}} 
\left(  \frac{1}{\sla{Q}-m}   \right) \nonumber    \\  &\times&  \frac{\partial}{\partial k_{\mu}}  \left( \frac{1}{\sla{Q}
-m}\right)\Big] \,, 
\end{eqnarray}
where we have used the identity $\left(\frac{\partial f(k+p) }{\partial p_{\alpha}} \right)_{p=0}
=\frac{\partial f(k) }{\partial k_{\alpha}}$.

We consider
\begin{eqnarray}
 \frac{\partial}{\partial k_{\mu}} 
\left(  \frac{1}{\sla{Q}-m}   \right)=\left( \frac{\partial Q_{\alpha}}{\partial k_{\mu}} 
\right)   \frac{1}{(\sla{Q}-m)^2}  \gamma^{\alpha}\,,
\end{eqnarray}
and after some algebra we arrive at
\begin{eqnarray}
\left(\frac{\partial^2\widetilde{\Pi}(p)}{\partial p_{\mu}\partial p_{\nu}}\right)_{p=0}=-
\int  \frac{d^4k}{(2\pi)^4}\left(\frac{\partial Q_{\alpha}}{\partial k_{\mu}} \right)\left(
  \frac{\partial Q_{\sigma}}{\partial k_{\nu}}  
\right) T^{\alpha \sigma}\,,
\end{eqnarray}
with
\begin{eqnarray}
T^{\alpha \sigma}=\frac{4}{(Q^2-m^2)^2} \eta^{\alpha \sigma}+ \frac{32m^2}
 {(Q^2-m^2)^4}Q^{\alpha} Q^{\sigma}\,.
\end{eqnarray}
By using the relations
\begin{eqnarray}\label{rela1}
&& \left(\frac{\partial Q_{\alpha}}{\partial k_{\mu}} \right)\left( 
 \frac{\partial Q^{\alpha}}{\partial k_{\nu}}  
\right)=\eta^{\mu \nu}-4g_2 n^{\mu} n^{\nu} (n\cdot Q)\,,  \nonumber \\
&& \left(\frac{\partial Q_{\alpha}}{\partial k_{\mu}} \right)  \left(\frac{\partial
 Q_{\sigma}}{\partial k_{\nu}} \right)  
Q^{\alpha}  Q^{\sigma}=\frac{1}{4} \left(\frac{\partial Q^2}{\partial 
k_{\mu}} \right)\left(\frac{\partial Q^2}{\partial k_{\nu}} \right)\,,
\end{eqnarray}
one obtains
\begin{eqnarray}\label{second-or}
 && \left(\frac{\partial^2\widetilde{\Pi}(p)}{\partial p_{\mu}\partial
  p_{\nu}}\right)_{p=0}=-4\int
  \frac{d^4k}{(2\pi)^4}  \left(   \frac{\eta^{\mu \nu}
  -4g_2 n^{\mu} n^{\nu} (n\cdot Q) }{(Q^2-m^2)^2} \right.   \nonumber \\ && \left. + 
\frac{2m^2}{(Q^2-m^2)^4}\left(\frac{\partial Q^2}
{\partial k_{\mu}}\right) \left(
\frac{\partial Q^2}{\partial k_{\nu}}\right)  \right)\,. 
\end{eqnarray}
Considering the tensors available  in our model, which are
the flat metric $\eta_{\mu \nu}$ and the preferred four-vector $n_{\mu}$
we can write
\begin{eqnarray}
\int \frac{d^4k}{(2\pi)^4} \frac{Q_{\mu}}{(Q^2-m^2)^2} =n_{\mu}S\,,
\end{eqnarray}
and 
\begin{eqnarray}\label{exp1}
&&\int \frac{d^4k}{(2\pi)^4} \frac{1}{(Q^2-m^2)^4}\left(\frac{\partial Q^2}
{\partial k_{\mu}}\right) \left(
\frac{\partial Q^2}{\partial k_{\nu}}\right)=\nonumber\\ &=&n^{\mu}n^{\nu}L+\eta^{\mu \nu}n^2M\,.
\end{eqnarray}
Now, consider the relation
\begin{eqnarray}
 \frac{\partial Q^2}{\partial k_{\mu}}=2\left(Q^{\mu}  -2g_2n^{\mu}(n\cdot Q)(n\cdot k)\right)\,,
\end{eqnarray}
 and multiplying it by $n_{\mu}n_{\nu}$ we arrive at
\begin{eqnarray}
&& \int \frac{d^4k}{(2\pi)^4} \frac{4(n\cdot Q)^2}{(Q^2-m^2)^4}  
\left(   1-4g_2n^2(n\cdot k) \right.  \nonumber \\ &+&\left. 4g_2^2(n^2)^2(n\cdot k)^2   \right) 
=
(n^2)^2(L+M)\,,
\end{eqnarray}
and by contracting with the metric $\eta_{\mu \nu}$
\begin{eqnarray}
&& \int \frac{d^4k}{(2\pi)^4} \frac{4}{(Q^2-m^2)^4}  \left(   Q^2-4g_2(n\cdot Q) ^2(n\cdot k) \right. 
\nonumber \\ &&  \left. +4g_2^2n^2(n\cdot Q) ^2(n\cdot k)^2   \right)=n^2(L+4M)\,.
\end{eqnarray}
Solving the algebraic equation we have
\begin{eqnarray}\label{L,M}
L&=&\frac{16}{3n^2}\int \frac{d^4k}{(2\pi)^4}   \frac{1}{(Q^2-m^2)^4}  
 \left(\frac{(n\cdot Q)^2}{n^2} - \frac{ Q^2}{4}  \right. 
\nonumber \\ &&  \left.  -3g_2(n\cdot Q)^2(n\cdot k)+3g_2^2
 (n\cdot Q)^2(n\cdot k)^2n^2   \right)\,,\nonumber
\\
M&=&\frac{4}{3n^2}\int \frac{d^4k}{(2\pi)^4}   \frac{1}{(Q^2-m^2)^4}  
 \left(  Q^2-\frac{ (n\cdot Q)^2}{n^2}     \right)\,.
\end{eqnarray}
A similar analysis gives
\begin{eqnarray}\label{intR}
S=\frac{1}{n^2}\int \frac{d^4k}{(2\pi)^4}   \frac{(n\cdot Q)}{(Q^2-m^2)^2}\,.
\end{eqnarray}
We find the second-order contribution
\begin{eqnarray}\label{secor}
 &&  \left(\frac{\partial^2\widetilde{\Pi}(p)}{\partial p_{\mu}\partial
  p_{\nu}}\right)_{p=0}p^{\mu} p^{\nu} =- \left(   4P+8m^2n^2M \right)p^2  
\nonumber \\ &&      +\left(16g_2n^2S-8m^2L\right)(n\cdot p)^2\,. 
\end{eqnarray}
Reorganizing this expression, we can write the correction to the scalar propagator up to second-order in $p$ 
as
\begin{eqnarray}\label{Pi}
\widetilde \Pi (p)&=&m^2 q_0+p^2q_1+(n\cdot p)^2q_n\,,
\end{eqnarray}
where
\begin{eqnarray}\label{defq_0}
q_0&=&\frac{K}{m^2}+2P\,,   \\ 
 q_1&=&-4 \left(   P+2m^2n^2M \right)\,,\nonumber
 \\ 
 q_n&=&8\left(2g_2n^2S-m^2L\right)\,.\nonumber
\end{eqnarray}
Finally one has
\begin{eqnarray}\label{q_0}
q_0&=&-\frac{i}{48\pi^2g_2^2m^2}
+\frac{i}{16\pi^2}
\left(2(3\gamma_E-1)-0.46    \right.  \\   && \left. -3\ln \left(\frac{g_2^2m^2}{4}\right)\right)  \nonumber \,,
\\ \label{q_1}
q_1&=&-\frac{ i}{12\pi^2}\left(-5+6\gamma_E  -0.46-3\ln\left(\frac{g_2^2m^2}{4}\right)     \right)  \,,
\\ \label{q_n}
q_n&=&\frac{i}{\pi^2}\,.
\end{eqnarray}
We provide details of the computation of 
$q_0$, $q_1$ and $q_n$ in the Appendix~\ref{AppendixA}.
The two-point function is finite and involves
a fine-tuning term proportional to $g_2^{-2}$. The Lee-Wick modes
improve the convergence of the theory such to make
the two-point function of the scalar field essentially UV finite and
involves the aether term~\cite{ouraether}.
\subsection{Fermion self-energy ${\Sigma}(p)$ }\label{subSec6}
Now we focus on the contribution of the fermion
self-energy graph depicted in Fig.~\ref{Fig5}, and recall that here
we consider $M=m$. 
\begin{figure}[H]
\centering
\includegraphics[width=0.3 \textwidth]{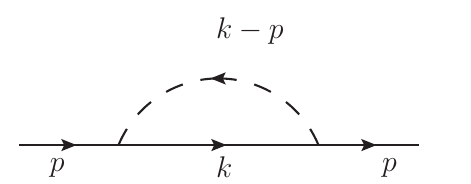}
\caption{\label{Fig5} The fermion self-energy graph.}
\end{figure}
The fermion self-energy graph is represented by the integral 
\begin{eqnarray}
i{\Sigma}(p)=g^2\int\frac{d^4k}{(2\pi)^4} \frac{\sla{Q}+m}
 {((k-p)^2-m^2) ( Q^2-m^2) }   \,.
\end{eqnarray}
To find it, let us consider a Taylor expansion of the first denominator term
up to second-order in $p$
and rewrite this contribution as
\begin{eqnarray}\label{self}
&&i{\Sigma}(p)\approx g^2\int \frac{d^4k}{(2\pi)^4} \left(
\frac{1}{k^2-m^2}+
\frac{2(k\cdot p)-p^2}{(k^2-m^2)^2   }+\right.\nonumber\\ &+&\left.
 \frac{4(k\cdot p)^2}{(k^2-m^2)^3}\right)\left(  \frac{\sla{Q}+m}
 { Q^2-m^2}   \right) +\mathcal O(p^3,n) \,.
\end{eqnarray}
With the notation $i{\Sigma}(p)=g^2I^{(0)}+g^2I^{(1)}+g^2I^{(2)}$, we introduce the 
the zeroth-order contribution
\begin{eqnarray}\label{I_0}
I^{(0)}=\int\frac{d^4k}{(2\pi)^4} 
\left(\frac{1}{k^2-m^2}\right)  \left(  \frac{\sla{Q}+m}
 { Q^2-m^2}   \right) \,,
\end{eqnarray}
the linear-order contribution
\begin{eqnarray}\label{I_1}
I^{(1)}=\int\frac{d^4k}{(2\pi)^4} 
\left( \frac{2(k\cdot p)}{(k^2-m^2)^2   } \right)  \left(  \frac{\sla{Q}+m}
 { Q^2-m^2}   \right) \,,
\end{eqnarray}
and the second-order contribution
\begin{eqnarray}\label{I_2}
I^{(2)}&=&\int\frac{d^4k}{(2\pi)^4} 
\left( \frac{4(k\cdot p)^2}{(k^2-m^2)^3} -\frac{p^2}{(k^2-m^2)^2   }\right)
 \left(  \frac{\sla{Q}+m}
 { Q^2-m^2}   \right) \,. \nonumber \\
\end{eqnarray}
\subsection{The gamma-matrix structure of $I^{(0)}$, $I^{(1)}$  $I^{(2)}$}
\subsubsection{Zeroth-order $I^{(0)}$}
Let us start with Eq.~(\ref{I_0}) and rewrite it as
\begin{eqnarray}\label{zeroth}
I^{(0)}&=&\gamma^{\mu}I_{\mu}^{(0)}+mf_0\,,
\end{eqnarray}
where we have defined 
 \begin{eqnarray}
I_{\mu}^{(0)}&=&\int\frac{d^4k}{(2\pi)^4} 
\frac{Q_{\mu} }{(k^2-m^2)(Q^2-m^2)} \,,
\end{eqnarray}
and
\begin{eqnarray}\label{f_0}
f_0&=& \int \frac{d^4k}{(2\pi)^4} 
\frac{1  }{(k^2-m^2)(Q^2-m^2)}\,.
\end{eqnarray}
From tensor analysis considerations one should have
\begin{eqnarray}\label{Q_mu}
\int\frac{d^4k}{(2\pi)^4} 
\frac{Q_{\mu} }{(k^2-m^2)(Q^2-m^2)} =n_{\mu}f_1^n\,.
\end{eqnarray}
Replacing the expression~(\ref{Q_mu}) in Eq.~(\ref{zeroth})
produces the zeroth-order contribution 
\begin{eqnarray}
I^{(0)}=\sla{n}f_1^n+mf_0\,,
\end{eqnarray}
with
\begin{eqnarray}\label{f_1^n}
f_1^n &=&\frac{1}{n^2} \int\frac{d^4k}{(2\pi)^4} 
\frac{(n\cdot Q) }{(k^2-m^2)(Q^2-m^2)}\,.
\end{eqnarray}
We carry out the calculations of $f_0$ and $f_1^n$
 following the lines given in Appendix~\ref{subAppendixB2}.
The first coefficient $f_0$ is naturally finite and the second one $f_1^n$ is divergent and
contains a large Lorentz-breaking correction term of the order of $g_2^{-1}$.
\subsubsection{Linear-order $I^{(1)}$}
The linear-order integral (\ref{I_1}) can be rewritten by introducing
\begin{eqnarray}\label{first-order}
I^{(1)}=2p^{\mu}  \gamma^{\nu} I_{\mu \nu}^{(1)}
+2mp^{\mu} I_{\mu}^{(1)}\,,
\end{eqnarray}
where
\begin{eqnarray}\label{Int1} 
 I_{\mu}^{(1)}&=&\int\frac{d^4k}{(2\pi)^4} 
\frac{k_{\mu} }{(k^2-m^2)^2(Q^2-m^2)}\,, \\  	\label{Int2}
 I_{\mu \nu}^{(1)}&=&\int\frac{d^4k}{(2\pi)^4} 
\frac{k_{\mu} Q_{\nu} }{(k^2-m^2)^2(Q^2-m^2)}\nonumber \,.
\end{eqnarray}
By considering 
\begin{eqnarray}
\int\frac{d^4k}{(2\pi)^4} 
\frac{k_{\mu} }{(k^2-m^2)^2(Q^2-m^2)}&=&n_{\mu}D \,,
\\
\int\frac{d^4k}{(2\pi)^4} 
\frac{k_{\mu} Q_{\nu} }{(k^2-m^2)^2(Q^2-m^2)}&=&n_{\mu}n_{\nu}B+n^2\eta_{\mu \nu} C\,,\nonumber 
\end{eqnarray}
and after some manipulations one finds
\begin{eqnarray}
D=\frac{1}{n^2}\int\frac{d^4k}{(2\pi)^4} 
\frac{(n\cdot k) }{(k^2-m^2)^2(Q^2-m^2)}\,,
\\
B=\frac{1}{3n^2}\int\frac{d^4k}{(2\pi)^4} 
\frac{\frac{4(n\cdot k)(n\cdot Q) }{n^2}- (k\cdot Q)  }{(k^2-m^2)^2(Q^2-m^2)}\,,\nonumber  \\
C=\frac{1}{3n^2}\int\frac{d^4k}{(2\pi)^4} 
\frac{  (k\cdot Q) - \frac{(n\cdot k)   (n\cdot Q)  }{n^2}}{(k^2-m^2)^2(Q^2-m^2)}\,.\nonumber 
\end{eqnarray}
Introducing the new notation we can write
\begin{eqnarray}
I^{(1)}=\sla{p}f_1+m(n\cdot p)f_2^n+\sla{n}(n\cdot p) f_3^n\,,
\end{eqnarray}
where
\begin{eqnarray}\label{f_1}
f_1&=&2n^2C \,,\\ \label{f_2^n}
f_2^n&=&2D \,,
\\ \label{f_3^n}
f_3^n&=&2B\,.
\end{eqnarray}
The terms $f_1$, $f_2^n$ and $f_3^n$ are convergent and
are explicitly calculated in Appendix~\ref{subAppendixB3}.
\subsubsection{Second-order $I^{(2)}$}
Following the same methodology the integral $I^{(2)}$  can be written as
\begin{eqnarray}\label{I^2}
I^{(2)}&=&-p^2\left(\gamma^{\mu}  { {I}}_{\mu}^{(2)}+m  {f_2 } \right)  +4p^{\mu}p^{\nu}\left(\gamma^{\alpha} 
 { { I}}_{\mu \nu \alpha}^{(2)}+m  { { I}}_{\mu \nu}^{(2)}\right)\,,
\end{eqnarray}
with
\begin{eqnarray}\label{tensors}
{  {I}}_{\mu }^{(2)}&=&\int\frac{d^4k}{(2\pi)^4} 
\frac{Q_{\mu}  }{(k^2-m^2)^2(Q^2-m^2)}\,,\nonumber\\ 
 {  {I}}_{\mu \nu}^{(2)}&=&\int\frac{d^4k}{(2\pi)^4} 
\frac{k_{\mu} k_{\nu} }{(k^2-m^2)^3(Q^2-m^2)}\,,\nonumber\\ 
{  {I}}_{\mu \nu\alpha}^{(2)}&=&\int\frac{d^4k}{(2\pi)^4} 
\frac{k_{\mu} k_{\nu} Q_{\alpha}}{(k^2-m^2)^3(Q^2-m^2)}\,,\nonumber
\\
f_2&=&\int\frac{d^4k}{(2\pi)^4} 
\frac{1 }{(k^2-m^2)^2(Q^2-m^2)}\,.
\end{eqnarray}
We have that ${  {I}}_{\mu \nu}^{(2)}$ and $f_2$ are of the order $g_2^2$ so they can be neglected.
For the remaining we define them with the following tensor structure 
\begin{eqnarray}
\int\frac{d^4k}{(2\pi)^4} 
\frac{Q_{\mu} }{(k^2-m^2)^2(Q^2-m^2)}&=&n_{\mu}\bar C\,,
\end{eqnarray}
\begin{eqnarray}
&& \int\frac{d^4k}{(2\pi)^4} 
\frac{k_{\mu}k_{\nu} Q_{\alpha} }{(k^2-m^2)^3(Q^2-m^2)}
=n_{\mu}n_{\nu}n_{\alpha}\bar D   \\    &&   +n^2\eta_{\mu \nu}
n_{\alpha} \bar E+n^2 \eta_{\nu \alpha}n_{\mu}\bar F \,.\nonumber
\end{eqnarray}
After some algebra we find
\begin{eqnarray}\label{int_LIV}
\bar C&=&\frac{1}{n^2}\int\frac{d^4k}{(2\pi)^4} 
\frac{  (n\cdot Q)   }{(k^2-m^2)^2(Q^2-m^2)},\nonumber 
\\
\bar E&=&\frac{1}{3(n^2)^2}\int\frac{d^4k}{(2\pi)^4} 
\frac{ \left(k^2-\frac{(n\cdot k)^2 }{n^2} \right)(n\cdot Q) }
{(k^2-m^2)^3(Q^2-m^2)}\,,\nonumber  \\
\bar D&=&\frac{1}{3(n^2)^2}\times\nonumber\\ &\times& 
\int\frac{d^4k}{(2\pi)^4} 
\frac{\left( \frac{5(n\cdot k)^2(n\cdot Q) }{n^2} -k^2(n\cdot Q)-  
 (n\cdot k)  (k\cdot Q) \right)}   {(k^2-m^2)^3(Q^2-m^2)},\, 
\nonumber \\
\bar F&=&-\frac{1}{3(n^2)^2}\int\frac{d^4k}{(2\pi)^4} 
\frac{ \left(\frac{(n\cdot k)^2(n\cdot Q) }{n^2} - (n\cdot k)  (k\cdot Q) \right)}   {(k^2-m^2)^3(Q^2-m^2)}.\, 
\end{eqnarray}
Replacing the expressions (\ref{tensors}) in (\ref{I^2}) and using the
 integrals (\ref{int_LIV}), up to linear-order in Lorentz violation 
and second-order in $p$, we arrive at
\begin{eqnarray}
I^{(2)}&=&p^2\sla{n}(-\bar C+4n^2\bar E)
+ 4n^2 (n\cdot p)\sla{p}  \bar F
+4(n\cdot p)^2\sla{n} \bar D\,.
\end{eqnarray}
We rewrite as
\begin{eqnarray}
I^{(2)}&=&p^2\sla{n}f_6^n
+  (n\cdot p)\sla{p}   f_4^n
+(n\cdot p)^2\sla{n}f_5^n\,,
\end{eqnarray}
with 
\begin{eqnarray}
  f_4^n&=&4n^2\bar F\,, \label{f_4}\\
f_5^n&=&4\bar D\,,  \label{f_5}\\
f_6^n&=&-\bar C+4n^2\bar E \label{f_6}\,,
\end{eqnarray}
where the explicit form of the relevant constants
 is given in the Appendix~\ref{subAppendixB4}.
\section{Mass renormalization}\label{Sec6}
\subsection{The on-shell subtraction scheme}\label{subSec61}
Having computed the one-loop correction to the scalar and fermion propagators, we 
now proceed to determine the pole mass in our Yukawa-like model. 
The radiative corrections
essentially involve extra terms such as 
 the scalar $(n \cdot p)$, the matrix $\sla{n}$ and their combinations.
 The theory is finite in the scalar sector
 and in the fermion sector
the only divergency appears in the term $f_1^n$. Since the mass corrections are finite in both sectors,
in some sense we are working in the scheme in which the finite parts of the renormalized
 mass correspond to the pole mass, and hence one can say that we are working in the on-shell subtraction
 scheme. Most of the ideas and method of calculation in the derivation of the pole mass,
 which are given below, are along the lines developed 
 in the work of Ref.~\cite{ralf}.

Let us remind the leading quantum corrections in the scalar sector 
\begin{eqnarray}
 i\Pi (p)&=&-2g^2m^2 q_0-2g^2p^2q_1-2g^2(n\cdot p)^2q_n\,,
\end{eqnarray}
where we have restored the constant $-2g^2$. In the fermion sector we have found
\begin{eqnarray}
i\Sigma_2&=&g^2\sla{n}f_1^n+g^2mf_0+g^2\sla{p}f_1+g^2m(n\cdot p)f_2^n
+g^2\sla{n}(n\cdot p) f_3^n\nonumber \\   &&+g^2p^2\sla{n}f_6^n
+ g^2 (n\cdot p)\sla{p}   f_4^n
+g^2(n\cdot p)^2\sla{n}f_5^n\,.
\end{eqnarray}
The coefficients $q$ and $f$ have been computed in the Appendices~\ref{AppendixA} 
and~\ref{AppendixB}.
\subsection{The scalar pole mass}\label{subSec62}
Let us start with the finite scalar Lagrangian 
\begin{eqnarray}
{\cal L}_{\phi}&=&\frac{1}{2}\partial_{\mu} \phi   \partial^{\mu}\phi 
-\frac 1 2M^2\phi^2\,,
\end{eqnarray}
in which no counterterms are needed to control ultraviolet divergencies.
The renormalized scalar two-point function is given by
\begin{eqnarray}\label{renPi}
(\Gamma_{\phi,R}^{(2)})^{-1}=p^2-M^2+\Pi_R(p)\,,
\end{eqnarray}
where
\begin{eqnarray}
\Pi_R(p)=p^2A_{\phi} +m^2 B_{\phi}+(n\cdot p)^2C_{\phi}\,,
\end{eqnarray}
where $A_{\phi} $, $B_{\phi}$ and $C_{\phi}$ are constants that can be deduced from 
the expressions~(\ref{q_0}),~(\ref{q_1}),~(\ref{q_n}) being
\begin{eqnarray}
iA_{\phi}&=&-2g^2q_1  \,,  \nonumber \\ 
iB_{\phi}&=&-2g^2q_0 \,,\nonumber\\
 iC_{\phi}&=&-2g^2q_n\,.
\end{eqnarray}
We define
\begin{eqnarray}\label{scalarbarP}
\bar P^2_{\phi}= p^2- { M}^2_{{\rm ph}}+\bar y(n\cdot p)^2\,.
\end{eqnarray}
in terms of the two unknown constants $M_{{\rm ph}}$ and $\bar y$,
Both constants can be determine
with the renormalization condition at $\bar P^2_{\phi}=0$
\begin{eqnarray}\label{scalarcond}
(\Gamma_{\phi,R}^{(2)})^{-1}(\bar P_{\phi}^2=0)=0\,.
\end{eqnarray}
Hence, from (\ref{renPi}) replacing the value of $p^2$ given in (\ref{scalarbarP})
and using the condition (\ref{scalarcond}), we arrive at the equation
\begin{eqnarray}
0&=&{ M}^2_{{\rm ph}}   -\bar y(n\cdot p)^2-M^2+  A_{\phi} 
\left({ M}^2_{{\rm ph}}   -\bar y(n\cdot p)^2\right)  \nonumber \\ &+&B_{\phi}m^2+C_{\phi}(n\cdot p)^2\,.
\end{eqnarray}
Due to the independence of each term, we find the two constants
\begin{eqnarray}
{ M}^2_{{\rm ph}} =\frac{(M^2-B_{\phi}m^2)}{(1+A_{\phi})}\,, \qquad \bar y =\frac{C_{\phi}}{1+A_{\phi}}\,,
\end{eqnarray}
and in consequence also the scalar pole mass $\bar P_{\phi}^2$. 
Substituting the above expressions in Eq.~(\ref{renPi}) and using 
\begin{eqnarray}
\frac{(\Gamma_{\phi,R}^{(2)})^{-1}(0)}{d\bar P^2_{\phi}}=Z_{\phi}^{-1}\,,
\end{eqnarray}
we identify 
the wave function normalization $Z_{\phi}^{-1}=1+A_{\phi}$, which in this case is finite.
\subsection{The fermion pole mass}\label{subSec63}
The fermionic Lagrangian written in terms of renormalized quantities reads
\begin{eqnarray}\label{yukawa1}
{\cal L}_{\psi}&=&i\bar{\psi} \ds  \psi-m \bar{\psi}  \psi   -\bar{\alpha}_R m\bar{\psi} \ns \psi
+g_2\bar{\psi} \sla{n} (n\cdot \partial)^2 \psi  \nonumber\\&-& \delta_{\bar{\alpha}}\bar \alpha_R m \bar{\psi} \sla{n} \psi\,,
\end{eqnarray}
where $\bar \alpha=Z_{\bar \alpha}\bar \alpha_R $ and  $Z_{\bar \alpha}=1+\delta_{\bar{\alpha}} $ and all other $Z_X$
equal to one.
The renormalized fermion two-point function is 
\begin{eqnarray}\label{two-point}
(\Gamma_{\psi,R}^{(2)})^{-1}=\sla{p}-m+\Sigma_R\,,
\end{eqnarray}
with
\begin{eqnarray}
\Sigma_R=\Sigma_2-\delta_{\bar{\alpha}} \bar \alpha_Rm \sla{n} +\mathcal O(g^4)\,.
\end{eqnarray}
We can write the finite contribution to
 the renormalized two-point function up to second-order in $p$ as
\begin{eqnarray}
\Sigma_R= \sla{p}A_{\psi}+mB_{\psi}+\sla{n}C_{\psi}\,,
\end{eqnarray}
where the explicit Lorentz violation coefficients at linear order in $g_2$, and 
coded in $\Sigma_2$, are
given by
\begin{eqnarray}
A_{\psi}&=&A^{(0)}+g_2A^{(1)}  (n\cdot p)\,,\nonumber\\
B_{\psi}&=&B^{(0)}+g_2B^{(1)}(n\cdot p)\,,\nonumber\\
C_{\psi}&=&\frac{C^{(0)}}{g_2}+C^{(1)}(n\cdot p)+g_2C^{(2)} (n\cdot p)^2+g_2 C^{(3)} p^2 \,.
\end{eqnarray}
We work in the minimal subtraction scheme and hence we fix the divergent term to be 
\begin{eqnarray}
\delta_{\bar{\alpha}}=\frac{g^2}{4\bar \alpha_R g_2m\pi^2\epsilon}\,,
\end{eqnarray}
in accordance with the correction term in (\ref{f_n^1}).
Within this scheme and according to the previous calculations we make the identification
\begin{eqnarray}
&&iA^{(0)}=g^2f_1\,,
\qquad
iA^{(1)}  =\frac{g^2f_4^n}{g_2}\,,\nonumber 
\qquad iB^{(0)}=g^2f_0\,,\nonumber 
\\ &&
iB^{(1)}=\frac{g^2f_2^n}{g_2}\,,\nonumber 
\qquad
iC^{(0)}=g_2(g^2f_1^n-\delta_{  \bar \alpha  } \bar \alpha_R)\,, \nonumber 
\\ &&
iC^{(1)}=g^2f_3^n\,, 
\qquad
iC^{(2)} =\frac{g^2f_5^n}{g_2}\,,  
\qquad
iC^{(3)} = \frac{g^2f_6^n}{g_2}\,.
\end{eqnarray}
To find the pole we consider the ansatz
\begin{eqnarray}\label{ansatz}
\bar P= \sla{p}-\bar m+\bar x \sla{n}\,,
\end{eqnarray}
where
\begin{eqnarray}
\bar m= m_{{\rm ph}}+g_2m_n (n\cdot p)\,,
\end{eqnarray}
and 
\begin{eqnarray}
\bar x= \frac{\bar x^{(0)}}{g_2}+\bar x^{(1)}(n\cdot p)+g_2\bar x^{(2)} (n\cdot p)^2+g_2 \bar x^{(3)} p^2\,.
\end{eqnarray}
We include the large Lorentz violating terms in the pole extraction process
which  has been included explicitly in $\bar x$.

Considering the condition for $\bar P$ to be a pole, that is to say, 
\begin{eqnarray}
(\Gamma_{\psi,R}^{(2)})^{-1}(\bar P=0)=\sla{p}-m+\Sigma_R=0\,,
\end{eqnarray}
provides us with
 the six terms $m_{ph}$, $m_{n}$, $\bar x^{(0)}$, $\bar x^{(1)}$, 
 $\bar x^{(2)}$, $\bar x^{(3)}$.
 We proceed as follows: we replace $\bar P$ in (\ref{two-point}),
  and arrive at
\begin{eqnarray}
(\Gamma_{\psi,R}^{(2)})^{-1}&=& \bar P+\bar m-\bar x \sla{n}-m+A(\bar P+\bar m-\bar x \sla{n})
 +Bm+C \sla{n}\,.\nonumber \\
\end{eqnarray}
With $\bar P=0$ we write
\begin{eqnarray}
&&(\Gamma_{\psi,R}^{(2)})^{-1}(\bar P=0)=0= m_{{\rm ph}}+
g_2m_n (n\cdot p)-\bar x \sla{n}-m\nonumber   \\ &+& \Big(A^{(0)}+gA^{(1)}
(n\cdot p) \Big) \Big(m_{{\rm ph}}+gm_n (n\cdot p)-\bar x \sla{n}\Big)  \nonumber 
\\&+&\left(B^{(0)}+g_2 B^{(1)}  (n\cdot p) \right)m+\left(\frac{1}
{g_2}C^{(0)}+C^{(1)}(n\cdot p)\right.  \nonumber  \\  &&   \left. + g_2 C^{(2)} (n\cdot p)^2+
g_2 C^{(3)}  p^2  \right) \sla{n}\,.
\end{eqnarray}
At lowest-order we have the first condition
\begin{eqnarray}
m_{{\rm ph}}-m+A^{(0)}m_{ph}+B^{(0)}m=0\,,
\end{eqnarray}
which has the solution
\begin{eqnarray}
m_{{\rm ph}}=m\frac{(1-B^{(0)})}{(1+A^{(0)})}\,.
\end{eqnarray}
Considering that all coefficients are independent, we have at linear-order for 
the factor accompanying $(n\cdot p)$:
\begin{eqnarray}
g_2 (n\cdot p)\left[m_n+A^{(0)}   m_n+A^{(1)}m_{{\rm ph}}+B^{(1)}m \right]=0\,,\nonumber \\
\end{eqnarray}
and consequently we have
\begin{eqnarray}
m_{n}=-m\left(\frac{(A^{(1)}(1-B^{(0)})}{(1+A^{(0)})^2}+ \frac{B^{(1)}}{1+A^{(0)}}\right)\,.
\end{eqnarray}
The next conditions determine $\bar x$ which follows from the equations
 with factor accompanying $\frac{1}{g_2} \sla{n}$, $(n\cdot p) \sla{n}$ ,
$g_2 (n\cdot p)^2 \sla{n}$, $g_2 p^2 \sla{n}$
respectively. Therefore, one arrives at
\begin{eqnarray}
\bar x^{(0)}&=&\frac{C^{(0)}}{1+A^{(0)}}\,,
\nonumber\\
\bar x^{(1)}&=&-\frac{A^{(1)}C^{(0)}}{(1+A^{(0)})^2}+\frac{C^{(1)}}{1+A^{(0)}}\,,\nonumber \\
\bar x^{(2)}&=&\frac{(A^{(1)})^2C^{(0)}}{(1+A^{(0)})^3}-
\frac{A^{(1)}C^{(1)}}{(1+A^{(0)})^2}+\frac{C^{(2)}}{1+A^{(0)}}\,,
\nonumber\\
\bar x^{(3)}&=&\frac{C^{(3)}}{1+A^{(0)}}\,.
\end{eqnarray}
This finishes the pole extraction in both sectors.

Let us perform an analysis of renormalization group equations and consider 
the beta function for the 
renormalized parameter $\bar \alpha_R$
\begin{eqnarray}
\beta(\bar \alpha_R)=\mu \frac{d}{d\mu} \bar \alpha_R\,.
\end{eqnarray}
Recall, in our theory the $\bar{\alpha}$ parameter is renormalized through the relation
\bea
\label{1}
\bar{\alpha}= \bar{\alpha}_R\left(1+\frac{ g^2}{4\bar{\alpha}_Rg_2m\pi^2\epsilon}\right)\,,
\eea
where in terms of the bare coupling $g_0$ one has in $d$ 
dimension $g_0=\mu^{\epsilon/2}g$. From the condition
\begin{eqnarray}
\mu \frac{d}{d\mu} \bar \alpha=0\,,
\end{eqnarray}
we find at leading order
\begin{eqnarray}
\beta(\bar \alpha_R)=  \frac{ g^2}{4 g_2m\pi^2}\,.
\end{eqnarray}
We use the renormalization group equation
\bea
\frac{d\bar{\alpha}_R}{dt}=\beta(\bar \alpha_R)\,,
\eea
which can be integrated as
\bea
\bar{\alpha}_R(t)=\bar{\alpha}_R(0)+\frac{g^2}{4g_2m\pi^2}t.
\eea
Hence we have no problems for low-energy domain but we can 
meet zero charge problem (Landau pole) at high energies
for $g_2<0$.
\section{Summary}\label{summary}
We considered the Myers-Pospelov-like higher-derivative extensions of the
Yukawa model which incorporates possible new physics from the Planck scale through
dimension five operators coupled to a preferred four vector $n^{\mu}$ which breaks the Lorentz symmetry.
We have selected a particular configuration of Lorentz symmetry violation
 in which the preferred four-vector $n^{\mu}$ is purely timelike. This choice
produces higher-order time-derivative terms and leads to extra solutions and new poles in the model.
We have found and identified the poles associated to standard 
solutions and those ones corresponding to negative-metric states or Lee-Wick solutions. 
Some of these poles move in the real axis, as in the usual Yukawa model. 
However, above a specific energy called the critical energy some solutions become complex 
introducing a extra difficulty for defining the consistent prescription for the contour of integration
 in the complex plane~\cite{Anselmi2}. To solve this problem we have 
 analyzed the motion of the poles in the complex plane.
In the scalar sector 
we have defined the usual Feynman integration contour which rounds the negative pole 
from below and the positive from above; the new poles that appear are located in the imaginary axis
and hence they do not present problems in this sense.
 However, in the fermion sector we have defined the prescription
with two considerations in mind:
first, the ability to recover the standard location of the poles relative to the real axis
when the Lorentz invariance violation is turned off. 
With this consideration we have imposed the minimal requirement on the contour $C_F^{(f)}$ 
to round the negative pole
$\omega_2$ from below and $\omega_1$ from above.
Second, to account for the other poles we have used the prescription 
putted forward in~\cite{QEDunitarity}, which has been shown to 
 lead to a unitary $S$-matrix in the regime of real solutions.
Using this Lee-Wick prescription, the three positive solutions $W_1$, $W_2$ and $\omega_1$
are defined to lie below the contour and the negative one $\omega_2$ above.
 At higher energies beyond the critical energy, we have called
for the following heuristic construction. We start at lower energies in which all the poles are real 
and where the basic requirements for consistency of the theory are satisfied. Next,
we begin to increase the energy at which complex solutions appear 
and define the new contour as the one obtained by 
 continuously deforming the first one in such a way to avoid
any crossing and singularity with the complex poles.

A central part of this work has been the study of mass renormalization
in the presence of both Lorentz-invariance violation and higher-order operators.
We have computed the one-loop corrections in the model and shown how they
lead to modifications in the renormalization procedure by pushing the pole mass to 
a sector involving other gamma matrices besides of $\sla{p}$.
The significance of our results, in particular, is based on the fact 
that this is one of the first works on the renormalization of higher-derivative Lorentz-breaking theories
within the context of modifications of asymptotic Hilbert space, whereas even without
 introducing the higher derivatives the renormalization issues in Lorentz-breaking
  theory were considered only in a few papers, in particular~\cite{One-loop,scarp,ba}. 
For other approaches in renormalization of higher-order Lorentz breaking theories, see~\cite{Anselmi1}.
We found that the theory is 
super-renormalizable (in principle, increasing the value of $N$ in the added kinetic operator
 $(n\cdot\partial)^N$, we can get a completely finite theory, 
the same result can be achieved through a supersymmetric extension of the theory, see~\cite{CMP}). 
The kinetic terms of the both fields are explicitly finite. As a by-product, 
we conclude that the aether-like terms originally introduced in \cite{ouraether, class} are generated 
within our studies both in scalar and spinor sectors, and they are essentially finite. It is interesting
 to note that in the spinor sector, the aether terms dominate in the low-energy limit in comparison 
 with the second-derivative term we introduced into the classical action.

At the same time, the large quantum corrections or, in other words, fine-tuning arise in our theory.
We have found in the scalar sector a large correction which in principle should be added to the 
usual fine tuning proportional to the square of the fermion mass. In consequence the effect of 
UV sensitivity of the scalar mass is maintained in our model. In the fermion sector we have that the large corrections
affect a term which do not involve any physical observable protecting the theory in this sector to a
genuine fine tuning. Actually, in our theory we have only one nontrivial counterterm, in the quadratic acton of the spinor.
 In a certain sense, this effect of large quantum corrections taking place 
in our theory resembles the effect of UV/IR mixing taking place in noncommutative field theories 
responsible for arising infrared singularities in a small noncommutativity limit \cite{Minw}.
Let us mention that in principle these large corrections
 are natural to expect since the higher-derivative extensions of the classical action can be 
treated as a kind of the higher-derivative regularization, hence, removing the regularization 
we return to singular results. We note that this effect is rather generic since the large quantum 
corrections are present even in the supersymmetric extensions of the Myers-Pospelov-like theories~\cite{CMP}.
The complete elimination 
of large quantum corrections could consist in employing
the extended supersymmetry which is well known to achieve complete finiteness
 of the field theory models as occurs for example for ${\cal N}=4$ super-Yang-Mills theory. Therefore, the 
 natural problem could consist in study of some Myers-Pospelov-like extensions of ${\cal N}=4$ 
 super-Yang-Mills theories. Another relevant problem consists in studying of the two-loop 
 approximation thus exhausting the possible divergences. We plan to carry out these studies in one of our next papers.
\begin{acknowledgements}
This work was partially supported by Conselho
Nacional de Desenvolvimento Cient\'{\i}fico e Tecnol\'{o}gico (CNPq). The work 
by A. Yu. P. has been supported by the
CNPq project \\ No. 303783/2015-0. CMR wants to thank the hospitality of the 
Universidade Federal da Paraiba
 (UFPB) and acknowledges support by FONDECYT grant 1140781. A. Yu. P. thanks the group 
 of \emph{F\'{\i}sica de Altas Energ\'{\i}as} of UBB for the hospitality. We also want to thank Markos Maniatis
 and York Schr\"oder
 for valuable comments and helpful discussions.
\end{acknowledgements}

\appendix

\section{The calculation of $q_0$, $q_1$, $q_n$ }\label{AppendixA}
We start to compute $q_0$ in (\ref{defq_0}), so we focus on $K$ and $P$ as given 
in Eqs.~(\ref{K}) and~(\ref{P}) and promote 
the integrals to $d$ dimensions 
and consider
\begin{eqnarray}\label{phasespace}
d^{d}k&=&\Omega_{d-1} |{\vec k}|^ {d-2} d|{\vec k}|\, dk_0\nonumber \\ &=&
\Omega_{d-1}E(E^2-m^2)^{\frac{d-3}{2}} dE\,dk_0\,, 
 \end{eqnarray}
where we have performed an integration in the angles 
producing the solid angle in $d-1$ dimensions 
 \begin{eqnarray}\label{d-solidangle}
 \Omega_{d-1}=\frac{2\pi^{(d-1)/2}}{\Gamma\left(\frac{d-1}{2}\right)}\,.
 \end{eqnarray}
From Eqs.~(\ref{K}) and~(\ref{P}), we note that 
 \begin{eqnarray}\label{K-P}
\frac{\partial K}{\partial m^2}=P\,.
 \end{eqnarray}
So we first compute $K$ and then derive $P$. 
To compute $K$ we consider the integral
\begin{eqnarray}
K&=&\mu^{4-d}  \Omega_{d-1} \int_m^{\infty} \frac{
 E  (E ^2-m^2)^{\frac{d-3}{2}}dE }
 {(2\pi)^d}  \nonumber  \\ &\times& \int _{C_F^{(f)}} \frac{dk_0}{(Q^2-m^2)}  \,.  
\end{eqnarray}
We rewrite the contour integrals making explicit 
their poles, and to compute them 
we use the method of residues and close
 the contour $C_F^{(f)}$ in the upper half plane as 
 shown in Fig.~\ref{Fig2}. 
We find 
\begin{eqnarray}\label{K_2}
&& \int _{C_F^{(f)}}  \frac{ dk_0}{g_2^2(k_0
-\omega_1 ) (k_0-\omega_2 ) (k_0-W_1 ) (k_0-W_2 )}   \nonumber \\ &=&
 - \frac{ \pi i}{  E  \sqrt{1 + 4 g_2 E }  } \,,
\end{eqnarray}
The next integral in the variable $E$ is direct, and gives at lowest order in $\epsilon=4-d$
the contribution
\begin{eqnarray}
K&=&  -\frac{i}{48\pi^2g_2^2}+  \frac{im^2}{48\pi^2}
\left(6\gamma_E-0.46 \right. \nonumber  \\  &-& \left. 3\ln \left(\frac{g_2^2m^2}{4}\right)\right)  \,.
\end{eqnarray}
From the relation \ref{K-P} one has
\begin{eqnarray}\label{res_p}
P&=& - \frac{i}{16\pi^2}+  \frac{i}{48\pi^2}
\left(6\gamma_E-0.46-3\ln \left(\frac{g_2^2m^2}{4}\right)\right)  \,.
\end{eqnarray}
It is simple to show that by combining the two contributions
through $q_0=K/m^2+2P$, produces
\begin{eqnarray}
q_0&=&-\frac{i}{48\pi^2g_2^2m^2}
+\frac{i}{16\pi^2}
\left(2(3\gamma_E-1)-0.46    \right. \nonumber  \\   && \left. -3\ln \left(\frac{g_2^2m^2}{4}\right)\right) \,.
\end{eqnarray}
The second-order contributions $q_1$ and $q_n$
are given in terms of $P$, $M$, $S$ and $L$ as shown in Eqs.~(\ref{defq_0}).
We begin with $M$, and since it is obtainable in four dimension, we
just set $d=4$, giving 
\begin{eqnarray}
M=-\frac{16\pi}{3} \int_m^{\infty} \frac{
 E  (E ^2-m^2)^{3/2}dE }
 {(2\pi)^4}    \int _{C_F^{(f)}}   \frac{  dk_{0}}{(Q^2-m^2)^4 }\,.
\end{eqnarray}
Solving the first integral gives 
\begin{eqnarray}
&&   \int _{C_F^{(f)}}   \frac{  dk_{0}}{g_2^8(k_0-\omega_1 )^4
 (k_0-\omega_2 )^4 (k_0-W_1 )^4 (k_0-W_2 ) ^4 } \nonumber   \\&=& \frac{\pi i \left(5+2g_2E\left(35+4g_2E
 \left(43+77g_2E\right)
 \right)\right)}{16E^7(1+4g_2E)^{7/2} }  \,,
\end{eqnarray}
and calculating then the $E$-integral, we arrive at
\begin{eqnarray}
M&=&-\frac{ i}{48\pi^2m^2}\,.
\end{eqnarray}
With the result of $P$ given in (\ref{res_p}), one has 
\begin{eqnarray}
q_1&=&-\frac{ i}{12\pi^2}\left(-5+6\gamma_E  -0.46-3\ln\left(\frac{g_2^2m^2}{4}\right)     \right)  \,.
\end{eqnarray}
We continue to compute $q_n$. From (\ref{intR}) we have
\begin{eqnarray}
S&=& \mu^{4-d} \Omega_{d-1}\int_m^{\infty} \frac{
 E  (E ^2-m^2)^{(d-3)/2}dE }
 {(2\pi)^d}  \\&\times&   \int _{C_F^{(f)}}    \frac{ k_{0}(1-g_2k_{0}) \;  dk_{0}}{g_2^4(k_0
-\omega_1 )^2 (k_0-\omega_2 )^2 (k_0-W_1 )^2 (k_0-W_2 ) ^2 }\,. \nonumber
\end{eqnarray}
In the same way, we arrive at
\begin{eqnarray}
S= \frac{ i}{16g_2\pi^2}+\frac{ig_2m^2}{16\pi^2}
\left(-4+\gamma_E-0.15-2\ln\left(\frac{g_2m}{2}\right) \right)\,.\nonumber \\
\end{eqnarray}
To compute $L$ given in the first equation (\ref{L,M}), 
\begin{eqnarray}
L&=&\frac{16}{3n^2}\int \frac{d^4k}{(2\pi)^4}   \frac{1}{(Q^2-m^2)^4}  
 \left(\frac{(n\cdot Q)^2}{n^2} - \frac{ Q^2}{4}\right. \nonumber  \\ && \left. -3g_2(n\cdot Q)^2(n\cdot k)+3g_2^2
 (n\cdot Q)^2(n\cdot k)^2n^2   \right)\,,
\end{eqnarray}
we note that
\begin{eqnarray}
L&=& L_2-M\,,
\end{eqnarray}
with
\begin{eqnarray}
L_2&=&16\pi \int_m^{\infty} \frac{
 E  (E ^2-m^2)^{1/2}dE }
 {(2\pi)^4}  \\ &\times&   \int _{C_F^{(f)}}   \frac{ k_{0}^2(1-g_2k_{0})^2 (1-2g_2 k_0)^2 \;  
dk_{0}}{g_2^8(k_0-\omega_1 )^4 
(k_0-\omega_2 )^4 (k_0-W_1 )^4 (k_0-W_2 ) ^4 }\,.\nonumber 
\end{eqnarray}
We find
\begin{eqnarray}
L_2&=&-\frac{ i}{48\pi^2m^2}\,,
\end{eqnarray}
and therefore $L=0$.
Considering the leading contributions for $q_n$ we obtain
\begin{eqnarray}
q_n&=&\frac{ i}{\pi^2}\,.
\end{eqnarray}
\section{The calculation of $\Sigma(p)$}\label{AppendixB}
\subsection{The general strategy}\label{subAppendixB1}
In this subsection, we describe the main steps to derive the basic integral that will appear
in the computation of the fermion self-energy correction~(\ref{self}). 
Let us start to consider the integral below where each order is labelled with $\alpha=1,2,3$ 
\begin{eqnarray}\label{Mint}
M_{(\alpha)}=\int \frac{d^4k}{(2\pi)^4}\frac{F(k_0,\vec k)}{(k^2-m^2)^{\alpha}(Q^2-m^2)}\,,
\end{eqnarray}
and where $F(k_0,\vec k)$ is an arbitrary function of $k_0$ and $\vec k$.

We promote the integrals to $d$ dimensions and recall the phase 
space measure given in~(\ref{phasespace}) in order to write
\begin{eqnarray}\label{M_alpha}
M_{(\alpha)}&=&\mu^{4-d}\Omega_{d-1}\int_m^{\infty} \frac{ dE}{(2\pi)^d} E (E^2-m^2)^{\frac{d-3}{2}}   
 \nonumber \\ &\times&    \int_{C_F^{(f)}}  \frac{F(k_0,\vec k)\; dk_0}{(k_0^2-E^2)^{\alpha}(
 (k_0- \bar{\alpha}m- g_2 k_0^2)^2-E^2)}\,.
\end{eqnarray}
To simplify the calculation, we will in the sequel everywhere approximate $(k_0- \bar{\alpha}m- 
g_2 k_0^2)^2-E^2)\simeq (k_0- g_2 k_0^2)^2-E^2)$ in all denominators, which will have 
very small modifications of the result since, first, $\bar{\alpha}$ is very small, second,
 it contributes only to subleading orders of results.
Working the denominator, we can rewrite the last contour integral as
\begin{eqnarray}
&&- \frac{1}{2Eg_2} \int_{C_F^{(f)}} dk_0 F(k_0,\vec k) \frac{1}{(k_0^2-E^2)^{\alpha}}  \nonumber \\ &\times&    \left (    \frac{1}{
  \left(\frac{k_0(-1+g_2k_0)+E}{g_2}\right) }-    \frac{1}{
\left(\frac{k_0(-1+g_2k_0)-E}{g_2}\right)  } \right)\,.
\end{eqnarray}
Now, using the Feynman parametrization
\begin{eqnarray}
\frac{1}{A^{\alpha}B^{\beta}}=\frac{\Gamma({\alpha}+\beta)}{
\Gamma({\alpha})\Gamma(\beta)}\int_0^1dx \frac{x^{{\alpha}-1}
 \left(1-x\right)^{\beta-1}}{\left(Ax+B\left(1-x\right)\right)^{{\alpha}+\beta}}\,,
\end{eqnarray}
and using $\beta =1$ we write the integral as
\begin{eqnarray}
 &&- \frac{\alpha }{2Eg_2}   \int _{C_F^{(f)}} dk_0 F(k_0,\vec k) 
   \int _0^1dx\nonumber  \\ &\times&
 \left [
  \frac{x^{\alpha-1}}{  \left[k_0^2  -\frac{k_0(1-x)}{g_2} -E^2 x
  +\frac{E(1-x)}{g_2}   \right]^{\alpha+1}}\right.\nonumber   \\  && \left.  -
   \frac{x^{\alpha-1}}{  \left[k_0^2  -\frac{k_0(1-x)}{g_2} -E^2 
   x-\frac{E(1-x)}{g_2}   \right]^{\alpha+1}}  \right] \,. 
\end{eqnarray}
Next, we carry out the change of variables 
$k_0^{\prime}=k_0-\frac{(1-x)}{2g_2}$
and perform a subsequent Wick rotation $k_0^{\prime}\to ik_{0E}^{\prime}$,
as a result, and dropping the tilde, we arrive at 
\begin{eqnarray}
&&    \frac{ i
(-1)^{\alpha} \alpha }{2Eg_2} \int _0^1dx \int _{-\infty}^{\infty}dk_{0E}  F\left(ik_{0E} 
+\frac{(1-x)}{2g_2} ,\vec k\right )\nonumber   \\ &\times&
\left[ \frac{x^{\alpha-1}}{  \left[k_{0E}^{2}  
+E^2 x-\frac{E(1-x)}{g_2}  +\frac{(1-x)^2}{4g_2^2} \right]^{\alpha+1}} 
\right.\nonumber   \\  && \left.    - \frac{x^{\alpha-1}}{  \left[k_{0E}^{2} 
+E^2 x+\frac{E(1-x)}{g_2}  +\frac{(1-x)^2}{4g_2^2} \right]^{\alpha+1}} \right]\,.
\end{eqnarray}
Replacing the above expression in~(\ref{M_alpha}) and again
changing variables by the rule $t=g_2E$ produces the final expression
\begin{eqnarray}\label{M_a}
M_{(\alpha)}&=& \frac{i 
(-1)^{\alpha}  \alpha \Omega_{d-1}\mu^{4-d}}{2}  
\nonumber \\  &\times& \int_{g_2m}^{\infty} \frac{ dt}{(2\pi)^d} \frac{1}{g_2^{d-1}} 
 (t^2-g_2^2m^2)^{\frac{(d-3)}{2}}  
l_{(\alpha)} \,, 
\end{eqnarray}
with 
\begin{eqnarray}\label{l_a}
l_{(\alpha)} &=& \int_{-\infty} ^{\infty} dk_{0E} \int _0^1dx F\left(ik_{0E} 
+\frac{(1-x)}{2g_2},\vec k \right ) \nonumber   \\  &\times&    \left[ \frac{x^{\alpha-1}}{  \left[k_{0E}^{2}  
+\frac{t^2}{g_2^2} x-\frac{t(1-x)}{g_2^2}  +\frac{(1-x)^2}{4g_2^2} \right]^{\alpha+1}} 
  \right.\nonumber\\ &-&\left. \frac{x^{\alpha-1}}{  \left[k_{0E}^{2}  
+\frac{t^2}{g_2^2} x+\frac{t(1-x)}{g_2^2}  +\frac{(1-x)^2}{4g_2^2} \right]^{\alpha+1}}  \right]\,,
\end{eqnarray}
which is the basic integral we need to solve in the next subsections.
\subsection{The zeroth-order terms $f_0$ and 
$f_1^n$ }\label{subAppendixB2}
We start to compute $f_0$ in $d$ dimensions which follows from~(\ref{f_0})
\begin{eqnarray}
f_0=\mu ^{4-d}\int \frac{d^dk}{(2\pi)^d} 
\frac{1  }{(k^2-m^2)(Q^2-m^2)}  \,.
\end{eqnarray}
We use the 
Eqs.~(\ref{M_a}) and~(\ref{l_a})
with the identification
 $\alpha=1$ and $F(k_0,\vec k)=1$, which yield
\begin{eqnarray}
f_0 = -\frac{i 
  \Omega_{d-1}\mu^{4-d}}{2} 
\int_{g_2m}^{\infty}
 \frac{ dt}{(2\pi)^d} \frac{1}{g_2^{d-1}} 
 (t^2-g_2^2m^2)^{\frac{(d-3)}{2}} l^{(1)}_1\,,\nonumber \\
\end{eqnarray}
with
\begin{eqnarray}\label{l_1^{(1)}}
l_1^{(1)}&=& 
\int_{-\infty} ^{\infty}dk_{0E}\int _0^1dx  \left[ \frac{1}{  \left[k_{0E}^{2}  
+\frac{t^2}{g_2^2} x-\frac{t(1-x)}{g_2^2}  +\frac{(1-x)^2}{4g_2^2}
 \right]^{2}}   \right.  \nonumber \\  &&  \left. -  \frac{1}{  \left[k_{0E}^{2}  +\frac{t^2}{g_2^2} x
+\frac{t(1-x)}{g_2^2}  +\frac{(1-x)^2}{4g_2^2} \right]^{2}}  \right]\,.
\end{eqnarray}
To proceed, we consider some useful integrals
\begin{eqnarray}\label{standint}
&&\int _{-\infty}^ {\infty} d k_{0E} \frac{1}{(k_{0E}^2+M^2)^{r}}=  \frac{\sqrt{\pi} 
}{\Gamma(r)}  \frac{\Gamma(r-\frac{1}{2})}{(M^2)^{r-\frac 1 2}}\,, 
\end{eqnarray}
\begin{eqnarray}\label{standint1}
&&\int _{-\infty}^ {\infty} d k_{0E}\frac{k_{0E}^2}{(k^2_{0E}+M^2)^{r}}= \frac{\sqrt{\pi} 
}{2\Gamma(r)}  \frac{\Gamma(r-\frac{3}{2})}{(M^2)^{r-\frac 3 2}}\,,  
\end{eqnarray}
\begin{eqnarray}\label{standint2}
&&\int _{-\infty}^ {\infty} d k_{0E}\frac{k_{0E}^4}{(k^2_{0E}+M^2)^{r}}= \frac{3\sqrt{\pi} 
}{4\Gamma(r)}  \frac{\Gamma(r-\frac{5}{2})}{(M^2)^{r-\frac 5 2}}\,.
\end{eqnarray}
By using the integral~(\ref{standint}) for $r=2$, we are able to 
solve the time integral in~(\ref{l_1^{(1)}}) arriving at
\begin{eqnarray}
l_1^{(1)}&=& \frac\pi 2 \int _0^1 dx \left[  \frac{ 1  }  {
(\frac{t^2}{g_2^2} x-\frac{t(1-x)}{g_2^2}  +\frac{(1-x)^2}{4g_2^2})^{3/2}   }   
  \right.  \nonumber \\  &&  \left. - \frac{ 1}  {
(\frac{t^2}{g_2^2} x+\frac{t(1-x)}{g_2^2}  +\frac{(1-x)^2}{4g_2^2})^{3/2}   }  \right].
\end{eqnarray}
Introducing the new variable $\varepsilon=t(1-x)$, we write
\begin{eqnarray}
l_1^{(1)}&=& \frac{g_2^3 \pi}{ 2} \int _0^1 dx \left[  \frac{1}  {
(t^2 x-\varepsilon  +\frac{(1-x)^2}{4})^{3/2}   }   \right.  \nonumber \\  &&  \left. -
 \frac{1}  {
(t^2 x+\varepsilon+\frac{(1-x)^2}{4})^{3/2}   } \right]\,.
\end{eqnarray}
We use the approximation
\begin{eqnarray} \label{approx}
 && \frac{ 1}  {
(t^2 x-\varepsilon  +\frac{(1-x)^2}{4})^{n/2}   }  -
 \frac{1}  {
(t^2 x+\varepsilon+\frac{(1-x)^2}{4})^{n/2}   } 
\nonumber \\ &=& \frac{ n\varepsilon}  {
(t^2 x +\frac{(1-x)^2}{4})^{n/2+1}   } 
+\mathcal O(\varepsilon^3)\,,
\end{eqnarray}
for $n=3$ followed by replacing $d=4$ in all explicitly finite terms, as 
a result, we arrive at the expression
\begin{eqnarray}
f_0= -\frac{3i 
    }{16\pi^2}  \int _0^1 dx\int_{g_2m}^{\infty} 
    dt \frac{t(1-x) (t^2-g_2^2m^2)^{\frac{d-3}{2}} }{(t^2 x+
     \frac{(1-x)^2}{4})^{5/2}} \,.
\end{eqnarray}
Integrating in $t$ and then in $x$ we arrive at the finite expression
\begin{eqnarray}
f_0&=&\frac{i}  {2\pi^2}  \left({\rm AppellF_1}
\left(-\frac{1}{2}, 1,1,\frac1 2 ,a,b\right)   \right.  \nonumber \\  &&
  \left. +{\rm AppellF_1}\left(\frac{1}{2}, 1,1,\frac3 2 ,a,b\right) \right )\,,
\end{eqnarray}
with the notation 
\begin{eqnarray}
\label{ab}
a&=&\frac{1}{1-2g_2^2m^2-2g_2m\sqrt{-1+g_2^2m^2}}\,,\nonumber \\
b&=&\frac{1}{1-2g_2^2m^2+2g_2m\sqrt{-1+g_2^2m^2}}.
\end{eqnarray} 
The function ${\rm AppellF_1}\left(a, b_1,b_2,c ,x_1,x_2\right)$ that appear above,
  is the hypergeometric function of two variables
defined by
\begin{eqnarray}
&&{\rm AppellF_1}
(a, b_1,b_2,c ,x_1,x_2) \nonumber \\ &=&\sum_{m=0}^{\infty} \sum_{n=0}^{\infty} \frac{ (a)_{m+n}
(b_1)_{m}(b_2)_{n}
} { m!\, n! \,(c)_{m+n}  } x_1^m\,x_2^n   \,, 
\end{eqnarray}
where $(q)_n=q(q+1)\dots (q+n-1)$, or equivalently
\begin{eqnarray}
(q)_n=\frac{\Gamma(q+n)}{\Gamma(q)}\,.
\end{eqnarray}

We continue with $f_1^n$, given in~(\ref{f_1^n}), and again promote to $d$ dimensions
\begin{eqnarray}
f_1^n =\mu ^{4-d}\int\frac{d^dk}{(2\pi)^d} 
\frac{k_0-\bar \alpha m- g_2 k_0^2 }{(k^2-m^2)(Q^2-m^2)} \,,
\end{eqnarray}
and using our Eqs.~(\ref{M_a}) and~(\ref{l_a}), 
with $\alpha=1$ and $F(k_0,\vec k)=k_0-\bar \alpha m-g_2 k_0^2$, we write
\begin{eqnarray}
f_1^n&=& -\frac{i 
\Omega_{d-1}\mu^{4-d}}{2}   \int_{gm}^{\infty}
\frac{ dt}{(2\pi)^d} \frac{1}{g_2^{d-1}} 
\nonumber \\ &\times& (t^2-g_2^2m^2)^{\frac{d-3}{2}}  l_2^{(1)}\,, 
\end{eqnarray}
with
\begin{eqnarray}
l_2^{(1)}&=& 
\int dk_{0E}\int _0^1dx \left( \frac{1-x}{2g_2}-  \bar \alpha m + g_2 
k_{0E}^2 -\frac{(1-x)^2}{4g_2}  \right)
   \nonumber \\  &\times&    \left[ \frac{1}{  \left[k_{0E}^{2}  
+\frac{t^2}{g_2^2} x-\frac{t(1-x)}{g_2^2}  +\frac{(1-x)^2}
{4g_2^2} \right]^{2}} \right. \nonumber\\ &-&\left.   \frac{1}
{  \left[k_{0E}^{2}  +\frac{t^2}{g_2^2} x+\frac{t(1-x)}{g_2^2} 
+\frac{(1-x)^2}{4g_2^2} \right]^{2}}  \right]\,.
\end{eqnarray}
Using the integrals~(\ref{standint}) and~(\ref{standint1})
we arrive at
\begin{eqnarray}\label{l_2^1}
l_2^{(1)}&=& \frac\pi 2 \int _0^1 dx \left[  
 \frac{ (\frac{1-x}{2g_2}- \bar \alpha m  -\frac{(1-x)^2}{4g_2}  )}  {
(\frac{t^2}{g_2^2} x-\frac{t(1-x)}{g_2^2}  +\frac{(1-x)^2}{4g_2^2})^{3/2}   }\nonumber  \right. \\ &-&  \left. 
 \frac{ (\frac{1-x}{2g_2}-  \bar \alpha m-\frac{(1-x)^2}{4g_2}  )}  {
(\frac{t^2}{g_2^2} x+\frac{t(1-x)}{g_2^2}   +\frac{(1-x)^2}{4g_2^2})^{3/2}   } \nonumber  \right. \\ &+&  \left.     \frac{g_2  }{
(\frac{t^2}{g_2^2} x-\frac{t(1-x)}{g^2}  +\frac{(1-x)^2}{4g_2^2})^{1/2}}  
\right. \nonumber\\ &-&\left. \frac{g_2  }{
(\frac{t^2}{g_2^2} x+\frac{t(1-x)}{g_2^2}  +\frac{(1-x)^2}{4g_2^2})^{1/2}}\right].
\end{eqnarray}
Just as in the previous calculation, we define $\varepsilon=t(1-x)$ 
and consider the approximation (\ref{approx}) with $n=3$ for the first 
and second term and $n=1$ for the third and fourth term in (\ref{l_2^1}).
Together with this, we replace $d=4$ in all the finite terms
and 
consider the identity 
$ \frac{1-x}{2}-\frac{(1-x)^2}{4}  =\frac{(1-x^2)}{4}$ to arrive at the simpler expression
\begin{eqnarray}
&& f_1^n= -\frac{i 
}{16\pi^2} \int _0^1 dx \frac{\mu^{\epsilon}}
{g_2^{d-3}} \int_{g_2m}^{\infty} dt\nonumber    \\  &\times&  \left(     \frac{  3 \left( \frac{(1-x^2)  }{4} + \bar \alpha m  \right)   t (1-x)
(t^2-g_2^2m^2)^{\frac{(d-3)}{2}} }{(t^2 x+ \frac{(1-x)^2}
{4})^{5/2}}   \nonumber \right.  \\  && \left. + \frac{t(1-x) (t^2-g_2^2m^2)^{\frac{(d-3)}{2}} }{(t^2 x+ 
 \frac{(1-x)^2}{4})^{3/2}} \right)\,.
\end{eqnarray}
Solving the $t$ integral and then the $x$ integrals and 
expanding in $\epsilon$, we are able to isolate the divergence, so to arrive at the final result
\begin{eqnarray}\label{f_n^1}
f_1^n&=&\frac{i }{4g_2\pi^2\epsilon}+ \frac{i}{8g_2\pi^2} \left[   (1-4\bar \alpha m)\right. \nonumber  \\ &\times& \left.  {\rm AppellF_1}
\left(-\frac{1}{2}, 1,1,\frac1 2 ,a,b\right)  \right.  \nonumber    \\  &+& \left.    (1-4\bar \alpha m){\rm AppellF_1}
\left(\frac{1}{2}, 1,1,\frac3 2 ,a,b\right)\nonumber \right.  \\ &+& \left.  \frac 13    {\rm AppellF_1}
\left(\frac{3}{2}, 1,1,\frac 5 2 ,a,b\right)   \nonumber \right. \\ &-& \left.  \frac 15 
{\rm AppellF_1}\left(\frac{5}{2}, 1,1,\frac 7 2 ,a,b\right)    
 \right]  \nonumber \\   &-&   \frac{i  }{8g_2\pi^2} \left[\gamma_E
-2\log(2g\mu)   \right.\nonumber  \\  &+&   \left.   {\rm PolyGamma}\left(0,\frac 32\right)\right]\,,  
\end{eqnarray}
where $a,b$ are given by (\ref{ab}) and ${\rm PolyGamma}\left(0,\frac 32\right)=0.036$.
\subsection{The linear-order terms $f_1$, $f_2^n$ and
$f_3^n$.}\label{subAppendixB3}
Now we compute the linear-order corrections given by the expressions
(\ref{f_1}), (\ref{f_2^n}) and (\ref{f_3^n}).
We start with 
\begin{eqnarray}
f_1&=&-\frac{2}{3}\mu^{4-d}\int \frac{d^dk}{(2\pi)^d} 
 \nonumber \\ &\times&  \frac{\vec k^2  }{(k_0^2-E^2)^2((k_0-  g_2 k_0^2)^2-E^2)}\,,
\end{eqnarray}
and rewrite it with $\alpha=2$ and $F(k_0,\vec k)=\vec k^2$, as
\begin{eqnarray}
f_1 &=&-\frac{2i\Omega_{d-1} \mu^{4-d} }{3}  \int_{g_2m}^{\infty} 
\frac{ dt}{(2\pi)^d} \frac{1}{g_2^{d-1}} \nonumber  \\  &\times& (t^2-g_2^2m^2)^{\frac{d-3}{2}}
l_1^{(2)}\,, 
\end{eqnarray}
where 
\begin{eqnarray}
l_1^{(2)}&=& 
\int dk_{0E}\int _0^1dx  \, \vec k^2\left[ \frac{x}{  
\left[k_{0E}^{2}  +\frac{t^2}{g_2^2} x-\frac{t(1-x)}{g_2^2}  
+\frac{(1-x)^2}{4g_2^2} \right]^{3}}   \right.  \nonumber    \\  && \left. 
 - \frac{x}{  \left[k_{0E}^{2}  +\frac{t^2}{g_2^2} x
 +\frac{t(1-x)}{g^2}  +\frac{(1-x)^2}{4g_2^2} \right]^{3}}  \right]\,.
\end{eqnarray}
Using (\ref{standint}) with $r=3$, we arrive at
\begin{eqnarray}
l_1^{(2)}&=& \frac{3\pi}{ 8} \int _0^1 dx  \,\vec k^2 \left[  \frac{ x  }  {
(\frac{t^2}{g_2^2} x-\frac{t(1-x)}{g_2^2}  +\frac{(1-x)^2}{4g_2^2})^{5/2}   }   
 \right.  \nonumber    \\  && \left.  -  \frac{ x}  {
(\frac{t^2}{g_2^2} x+\frac{t(1-x)}{g_2^2}  +\frac{(1-x)^2}{4g_2^2})^{5/2}   }  \right].
\end{eqnarray}
With $\varepsilon=t(1-x)$, $\vec k^2=\frac{1}{g_2^2}(t^2-g_2^2m^2)$ and 
using the approximation (\ref{approx}) gives
\begin{eqnarray}
f_1&=& -\frac{5i}
{16\pi^2}   \int _0^1 dx \frac{\mu^{\epsilon}}{g_2^{d-4}} \nonumber \\ &\times&
\left( \int_{g_2m}^{\infty} dt \frac{t(1-x) x (t^2-g_2^2m^2)
^{\frac{d-1}{2}} }{(t^2 x+ \frac{(1-x)^2}{4})^{7/2}} \right)\,. 
\end{eqnarray}
Again, performing the $t$ integral followed by the $x$ integral, we arrive at 
\begin{eqnarray}\label{calcf_1}
f_1&=&\frac{i}{2\pi^2}   \left[{\rm AppellF_1}\left(-\frac{1}{2}, 1,1,\frac1 2 ,a,b\right) 
+ \right.\nonumber\\ &+&\left.
{\rm AppellF_1}\left(\frac{1}{2}, 1,1,\frac3 2 ,a,b\right) \right ]\,,
\end{eqnarray}
with $a,b$ are given by (\ref{ab}). 

Now we consider
\begin{eqnarray}
f_2^n=2\mu^{4-d}\int\frac{d^dk}{(2\pi)^d}  
 \frac{ k_0  }{(k_0^2-E^2)^2((k_0-  g_2 k_0^2)^2-E^2)}\,,
\end{eqnarray}
which we rewrite as
\begin{eqnarray}
f_2^n &=&2i\Omega_{d-1} \mu^{4-d}  
 \int_{g_2m}^{\infty} \frac{ dt}{(2\pi)^d} \frac{1}{g_2^{d-1}} 
\nonumber \\ &\times& (t^2-g_2^2m^2)^{\frac{d-3}{2}} l_2^{(2)}\,.
\end{eqnarray}
After the same change of variables in $l_2^{(2)}$,
a linear term in $k_0^{\prime}$ vanishes and we are left with 
\begin{eqnarray}
l_2^{(2)}&=& \int dk_{0E}
\int _0^1dx  \frac{(1-x)}{2g_2}   \nonumber \\   &\times&   \int  \left[ \frac{x}{ 
\left[k_{0E}^{2}  +\frac{t^2}{g_2^2} x-\frac{t(1-x)}{g_2^2}  
+\frac{(1-x)^2}{4g_2^2} \right]^{3}}     \right.  \nonumber    \\  &-& \left.     \frac{x}{  \left[k_{0E}^{2}  +\frac{t^2}{g_2^2} x
 +\frac{t(1-x)}{g_2^2}  +\frac{(1-x)^2}{4g_2^2} \right]^{3}}  \right].
\end{eqnarray}
Using (\ref{standint}),
we have
\begin{eqnarray}
l_2^{(2)}&=&\frac{3\pi}{ 8} \int _0^1 dx \frac{(1-x)}{2g_2} \left[  \frac{ x  }  {
(\frac{t^2}{g_2^2} x-\frac{t(1-x)}{g_2^2}  +\frac{(1-x)^2}{4g_2^2})^{5/2}   }   
 \right.  \nonumber    \\  && \left. - \frac{ x}  {
(\frac{t^2}{g_2^2} x+\frac{t(1-x)}{g_2^2}  +\frac{(1-x)^2}{4g_2^2})^{5/2}   }  \right].
\end{eqnarray}
Again, we define $\varepsilon=t(1-x)$ and
consider again the approximation (\ref{approx})
and replace $d=4$ in all the finite terms. We arrive at the simpler expression
\begin{eqnarray}
f_2^n&=& \frac{15i}{32\pi^2}   \int _0^1 dx \frac{\mu^{\epsilon}}{g_2^{d-5}}  
\nonumber \\   &\times&\left( \int_{g_2m}^{\infty} dt \frac{(1-x)^2t x (t^2-g_2^2m^2)
^{\frac{(d-3)}{2}} }{(t^2 x+ \frac{(1-x)^2}{4})^{7/2}} \right)\,.
\end{eqnarray}
Integrating in $t$ and then in $x$ at lowest order, we arrive at 
\begin{eqnarray}\label{calcf_2^n}
f_2^n&=&\frac{2ig_2}{\pi^2}  \left[ {\rm AppellF_1}\left(\frac{1}{2}, 1,1,
\frac3 2 ,a,b\right)  \right ]\,, 
\end{eqnarray}
with the same $a,b$ given by (\ref{ab}).

To find $f_3^n$ we take advantage of having already calculated the piece $f_1$ and by considering 
$f_3^n=2B$
\begin{eqnarray} 
f_3^n&=&\frac{2}{(n^2)^2}\int \frac{d^4k}{(2\pi)^4} \frac{(n\cdot k)
(n\cdot Q)}{(k^2-m^2)^2(Q^2-m^2)}-f_1  \nonumber \\ & \equiv &T_1+T_2-f_1\,,
\end{eqnarray}
we focus on the pieces
\begin{eqnarray}
T_1=  \mu^{4-d}\int \frac{d^dk}{(2\pi)^d}
 \frac{k_0^2}{(k^2-m^2)^2(Q^2-m^2)}\,,
\end{eqnarray}
and 
\begin{eqnarray}
T_2=-g_2\mu^{4-d}\int d^dk\frac{k_0^3}{(k^2-m^2)^2(Q^2-m^2)}\,.
\end{eqnarray}
where we are considering $\bar \alpha$ small.

Following the same technique we find 
\begin{eqnarray}
T_1&=&-\frac{i}{4\pi^2} \left[{\rm AppellF_1}\left(\frac{1}{2}, 1,1,\frac3 2 ,a,b\right) 
 \right.      \\  && \left.  -\frac{1}{3}{\rm AppellF_1}\left(
\frac{3}{2}, 1,1,\frac5 2 ,a,b\right) \right ]   \nonumber \\ &+&
\frac{i}{2\pi^2}  \left[{\rm AppellF_1}
\left(\frac{1}{2}, 2,2,\frac3 2 ,a,b\right)  \right.  \nonumber    \\  && \left. 
- {\rm AppellF_1}\left(\frac{3}{2}, 2,2,
 \frac5 2 ,a,b\right)  \right.  \nonumber    \\  && \left. +\frac 3 5{\rm AppellF_1}
 \left(\frac{5}{2}, 2,2,\frac7 2 ,a,b\right)   \right.\nonumber   \\&-&\left.   \frac{1}{7}
 {\rm AppellF_1}\left(\frac{7}{2}, 2,2,\frac9 2 ,a,b\right) \right ]  \,,\nonumber
\end{eqnarray}
and 
\begin{eqnarray}
T_2&=&\frac{3i  } {8\pi^2}   \left[1-\frac{4g^2m^2}{3}{\rm 
AppellF_1}\left(\frac{3}{2}, 1,1,\frac5 2 ,a,b\right) \right ]\nonumber \\  &&
-\frac{i }{4\pi^2 }  \left[{\rm AppellF_1}\left(\frac{1}{2}, 2,2,\frac3 2 ,a,b\right)\right.  \nonumber    \\  && \left. -\frac 4 3
{\rm AppellF_1}\left(\frac{3}{2}, 2,2,\frac5 2 ,a,b\right) \right.
\nonumber \\ &+& \left. \frac{6}{5}{\rm 
AppellF_1}\left(\frac{5}{2}, 2,2,\frac7 2 ,a,b\right)\right.  \nonumber    \\  && \left. -\frac 4 7{\rm AppellF_1}
\left(\frac{7}{2}, 2,2,\frac9 2 ,a,b\right) \right.  \nonumber    \\  && \left.+\frac{1}{9}
{\rm AppellF_1}\left(\frac{9}{2}, 2,2,\frac{11 }{2} ,a,b\right) \right ]\,,
\end{eqnarray}
and therefore
\begin{eqnarray}
f_3^n&=&\frac{3i  } {8\pi^2}  
+\frac{i }{2\pi^2 }  \left[-{\rm AppellF_1}\left(-\frac{1}{2}, 1,1,\frac3 2 ,a,b\right) \right.   \nonumber    \\  &-& \left.
-\frac{3}{2}{\rm AppellF_1}\left(\frac{1}{2}, 1,1,\frac3 2 ,a,b\right)   \right.  \nonumber    \\  &+& \left. \frac 1 6
{\rm AppellF_1}\left(\frac{3}{2}, 1,1,\frac5 2 ,a,b\right) \right.
\nonumber \\ &+& \left. \frac{1}{2}{\rm 
AppellF_1}\left(\frac{1}{2}, 2,2,\frac3 2 ,a,b\right) \nonumber \right.  \\ &-& \left.  \frac 2 3{\rm AppellF_1}
\left(\frac{3}{2}, 2,2,\frac5 2 ,a,b\right) \right.  \nonumber    \\  &+& \left.\frac{1}{7}
{\rm AppellF_1}\left(\frac{7}{2}, 2,2,\frac{9 }{2} ,a,b\right) \right. \nonumber   \\
&-& \left.        \frac{1}{18}
{\rm AppellF_1}\left( \frac{9}{2}, 2,2,\frac{11 }{2} ,a,b\right)    \right]\,. 
\end{eqnarray}
with $a,b$ given by (\ref{ab}).
\subsection{The second-order terms $f_4^n$, $f_5^n$, $f_6^n$.}\label{subAppendixB4}
Here we impose the further simplification $\bar \alpha$
small which eventually may contribute to finite terms in the numerators of the terms below.
For the second-order contributions we start with $f_4^n$ in Eq.(\ref{f_4}).  
and consider
\begin{eqnarray}\label{barF}
\bar F=-\frac{1}{3}\mu^{4-d}\int\frac{d^dk}{(2\pi)^d} 
\frac{   \vec k^2 k_0}{(k^2-m^2)^3(Q^2-m^2)}\,.
\end{eqnarray}
We write
\begin{eqnarray}
\bar F&=&\frac{i}{2}\Omega_{d-1} \mu^{4-d}   \int_{g_2m}^{\infty} \frac{ dt}{(2\pi)^d} \frac{1}{g_2^{d-1}}
\nonumber \\ &\times& (t^2-g_2^2m^2)^{\frac{d-3}{2}} l_1^{(3)}\,, 
\end{eqnarray}
where 
\begin{eqnarray}
l_1^{(3)}&=& 
\int dk_{0E}  \left(  \vec k^2\frac{(1-x)}{2g_2}  \right) \nonumber\\
&\times& \left[ \int _0^1dx\frac{x^2}{  \left[k_{0E}^{2}  +\frac{t^2}{g_2^2} x
-\frac{t(1-x)}{g_2^2}  +\frac{(1-x)^2}{4g_2^2} \right]^{4}}   \right.  \nonumber    \\  && \left.
- \int _0^1dx\frac{x^2}{  \left[k_{0E}^{2}  +\frac{t^2}{g_2^2} x+\frac{t(1-x)}{g_2^2}  
+\frac{(1-x)^2}{4g_2^2} \right]^{4}}  \right]\,.  
\end{eqnarray}
Using (\ref{standint}), we arrive at
\begin{eqnarray}
l_1^{(3)}&=&\frac{5\pi}{ 16} \int _0^1 dx(\vec k^2) \frac{(1-x)}{2g_2} \left[  \frac{ x ^2 }  {
(\frac{t^2}{g_2^2} x-\frac{t(1-x)}{g_2^2}  +\frac{(1-x)^2}{4g_2^2})^{7/2}   }   \right.  \nonumber    \\  && \left.
- \frac{ x^2}  {
(\frac{t^2}{g_2^2} x+\frac{t(1-x)}{g_2^2}  +\frac{(1-x)^2}{4g_2^2})^{7/2}   }  \right]\,.
\end{eqnarray}
We again use $\varepsilon=t(1-x)$ and write
\begin{eqnarray}
l_1^{(3)}&=&\frac{5\pi g_2^6}{ 32} \int _0^1 dx \left(\frac{t^2-g_2^2m^2}{g_2^2}\right)  \left[  \frac{(1-x)x^2}  {
(t^2 x-\varepsilon  +\frac{(1-x)^2}{4})^{7/2}   }\right.  \nonumber    \\  && \left. - \frac{(1-x) x^2}  {
(t^2 x+\varepsilon+\frac{(1-x)^2}{4})^{7/2}   } \right]\,.
\end{eqnarray}
We consider the approximation (\ref{approx}) and
replace $d=4$ in all the finite terms. We arrive at 
\begin{eqnarray}
\bar F&=& \frac{35i 
    }{256\pi^2}   \int _0^1 dx \frac{\mu^{\epsilon}}{g_2^{d-5}}  \nonumber \\ &\times& \left( \int_{gm}^{\infty} dt 
   \frac{(1-x)^2t x^2 (t^2-g_2^2m^2)^{\frac{(d-1)}{2}} }{(t^2 x+ \frac{(1-x)^2}{4})^{9/2}} \right)\,.
\end{eqnarray}
Integrating in $t$ and then in $x$ and considering $f_4^n=4\bar F$
we have
\begin{eqnarray}
f_4^n&=&\frac{ig_2}{4\pi^2}\left[{\rm AppellF_1}
\left(\frac{1}{2}, 1,1,\frac3 2 ,a,b\right) \right ]
\,,
\end{eqnarray}
with $a,b$ given by (\ref{ab}).

Now we compute  $f_5^n$ and focus on $\bar D$. From the integrals
(\ref{int_LIV}) one can show that
\begin{eqnarray}
\bar {D}&=&\frac{1}{3}\int\frac{d^4k}{(2\pi)^4} 
\frac{(3k_0^3-3g_2k_0^4+k_0\vec k^2-g_2k_0^2\vec k^2)}
{(k^2-m^2)^3((k_0-  g_2 k_0^2)^2-E^2)} \nonumber \\ &-& \frac{f_4^n}{4}\,, 
\end{eqnarray}
and write 
\begin{eqnarray}
f_5^n=P-T-Y-2f_4^n\,,
\end{eqnarray}
by considering the definitions of the $d$ dimensions integrals
\begin{eqnarray}\label{secorint}
P&=&4\mu^{4-d}\int\frac{d^dk}{(2\pi)^d} 
\frac{k_0^3}{(k^2-m^2)^3((k_0-  g_2 k_0^2)^2-E^2)} \,,\nonumber
\\
T&=&4g_2\mu^{4-d}\int\frac{d^dk}{(2\pi)^d} 
\frac{k_0^4    }{(k^2-m^2)^3((k_0-  g_2 k_0^2)^2-E^2)}\,,\nonumber\\  
Y&=&4g_2\frac{\mu^{4-d}}{3}\int\frac{d^dk}{(2\pi)^d} \nonumber \\ &\times&
\frac{  \vec k^2k_0^2 }{(k^2-m^2)^3((k_0-  g_2 k_0^2)^2-E^2)}  \,.
\end{eqnarray}
Let us compute $P$, with $\alpha=3$, we have
\begin{eqnarray}
P&=&-\frac{3i}{2}\Omega_{d-1} \mu^{\epsilon}\nonumber  \\ &\times&  \int_{gm}^{\infty} \frac{ dt}{(2\pi)^d}
 \frac{1}{g_2^{(d-1)}} (t^2-g_2^2m^2)^{\frac{(d-3)}{2}} l_2^{(3)}\,, 
\end{eqnarray}
with
\begin{eqnarray}
l_2^{(3)}&=& 
\int dk_{0E}  \left( -3 k_{0E}^2\frac{(1-x)}{2g_2} +\frac{(1-x)^3}{8g_2^3}  \right) \\
&\times& \left[ \int _0^1dx\frac{x^2}{  \left[k_{0E}^{2}  +\frac{t^2}{g_2^2} x
-\frac{t(1-x)}{g_2^2}  +\frac{(1-x)^2}{4g_2^2} \right]^{4}}
\right.  \nonumber    \\  && \left. - \int _0^1dx\frac{x^2}{  \left[k_{0E}^{2}  
+\frac{t^2}{g_2^2} x+\frac{t(1-x)}{g_2^2}  +\frac{(1-x)^2}{4g_2^2} \right]^{4}}  \right]\,.\nonumber 
\end{eqnarray}
After some algebra we find at lowest order
\begin{eqnarray}
P&=&\frac{ig_2 }{\pi^2}  \left[{\rm AppellF_1}\left(\frac{3}{2}, 1,1,\frac5 2 ,a,b\right) \right.  \nonumber 
   \\  && \left.-\frac 4 3{\rm AppellF_1}\left(\frac{3}{2}, 3,3,\frac5 2 ,a,b\right) 
\right.  \nonumber    \\  && \left. +\frac {16}{ 5}{\rm AppellF_1}
\left(\frac{5}{2}, 3,3,\frac7 2 ,a,b\right) \right. \nonumber \\ &-&\left.\frac{24}{7}
{\rm AppellF_1}\left(\frac{7}{2}, 3,3,\frac9 2 ,a,b\right)
\right.  \nonumber    \\  && \left.  +\frac {16}{ 9}{\rm AppellF_1}\left(\frac{9}{2}, 3,3,\frac{11}{ 2} ,a,b\right)
 \right.  \nonumber    \\  && \left.-\frac{4}{11}{\rm AppellF_1}\left(\frac{11}{2}, 3,3,\frac{13 }{2} ,a,b\right) \right ]\,.
\end{eqnarray}
In the same way we find
\begin{eqnarray}
T&=&-\frac{i g_2 }{\pi^2} \left[\frac{1}{4}{\rm AppellF_1}\left(\frac{3}{2}, 1,1,\frac5 2 ,a,b\right) 
+  \right.\nonumber\\ &+&\left.  \frac 2 3{\rm AppellF_1}\left(\frac{3}{2}, 3,3,\frac5 2 ,a,b\right) 
\right.  \nonumber    \\  && \left. -2{\rm AppellF_1}\left
(\frac{10}{2}, 3,3,\frac7 2 ,a,b\right) \right.\nonumber\\ &+&\left. \frac{20}{7}{\rm AppellF_1}
\left(\frac{7}{2}, 3,3,\frac9 2 ,a,b\right)\right.  \nonumber    \\  && \left. -\frac {20}{9} {\rm AppellF_1}
\left(\frac{9}{2}, 3,3,\frac{11}{ 2} ,a,b\right) + \right.\nonumber\\ &+&\left.\frac{10}{11}{\rm AppellF_1}\left(\frac{11}{2}, 3,3,\frac{13}{2}  ,a,b\right)
  \right. 
\nonumber \\ &&  \left.    -\frac {2}{13} {\rm AppellF_1}\left(\frac{13}{2}, 3,3,\frac{15}{ 2} ,a,b\right) \right ]
\nonumber \\ &&+\frac{ig_2}{2\pi^2 }\left[2{\rm AppellF_1}\left(\frac{3}{2}, 2,2,\frac5 2 ,a,b\right) -\right. \nonumber\\ &-& \left.\frac {18}
{ 5}{\rm AppellF_1}\left(\frac{5}{2}, 2,2,\frac7 2 ,a,b\right) \right.
\nonumber\\ &+&\left.  \frac{18}{7}{\rm AppellF_1}\left(\frac{7}{2}, 2,2,\frac9 2 ,a,b\right) \right.
  \nonumber    \\  && \left.-\frac {1}{3} {\rm AppellF_1}\left(\frac{1}{3}, 2,2,\frac{11}{ 2} ,a,b\right) \right ]\,,
\end{eqnarray}
and
\begin{eqnarray}
Y&=&\frac{ig_2}{\pi^2}\left[{\rm AppellF_1}\left(\frac{1}{2}, 1,1,\frac3 2 ,a,b\right)  \nonumber  \right. \\& -&  \left. \frac{1}{3}
{\rm AppellF_1}\left(\frac{3}{2}, 1,1,\frac5 2 ,a,b\right) \right ]\nonumber 
\\&-&\frac{ig_2}{\pi^2 }\left[2{\rm AppellF_1}(\frac{1}{2}, 2,2,\frac3 2 ,a,b)   \nonumber  \right. \\& -&  \left.   2{\rm AppellF_1}
\left(\frac{3}{2}, 2,2,\frac5 2 ,a,b\right)\right.  \nonumber    \\  &+& \left.  \frac{6}{5}{\rm AppellF_1}\left(\frac{5}{2}, 2,2,\frac7 2 ,a,b\right) \right.
\nonumber\\ &-&\left. \frac {2}{7} {\rm AppellF_1}\left(\frac{7}{2}, 2,2,\frac{9}{ 2} ,a,b\right) \right ]\,,
\end{eqnarray}
with $a,b$ are given by (\ref{ab}). To find $f_5^n$ we just have to add the above contributions.

To compute $f_6^n=-\bar C+4\bar E$ we need the two pieces $\bar C$ and $\bar E$. We have 
 \begin{eqnarray}
\bar C=\frac{f_2^n}{2}-g_2T_1\,,
\end{eqnarray}
 and from (\ref{int_LIV})
 \begin{eqnarray}
\bar E=\frac{1}{3}\int\frac{d^4k}{(2\pi)^4} 
\frac{ -\vec k^2k_0+g_2\vec k^2k_0^2   }{(k^2-m^2)^3(Q^2-m^2)} \,,
\end{eqnarray}
which can be given in terms of coefficients we have computed before, from (\ref{secorint}), (\ref{barF})
and $f_4^n=4\bar F$,
as
\begin{eqnarray}
\bar E=\frac{Y}{4}+\frac{f_4^n}{4}\,.
\end{eqnarray}
Finally we can write
 \begin{eqnarray}
f_6^n=-\frac{f_2^n}{2}+g_2T_1+Y+f_4^n\,,
\end{eqnarray}
which can be derived with the expressions we have calculated before.
\section{Cutoff regularization}\label{AppendixC}
Here we consider the cut-off regularization scheme for some integrals that appear 
to be finite using dimensional regularization. 

We begin to focus on the integral $I_1$ of Eq. (\ref{Iuno}) with a momentum cutoff $\Lambda$. The integral is
\begin{eqnarray}
 I_1(\Lambda)=\int _{z_0}^{\Lambda}
dz \,  \frac{  \sqrt{ z^2-z_0^2 }  }{\sqrt{z+1}}   \,.
\end{eqnarray}
 After a straightforward calculation we get for $\Lambda \to \infty$
\begin{eqnarray}
I_1(\Lambda)&=&\frac{2}{3}\Lambda^{3/2}- \Lambda^{1/2}+ 
 \frac{4}{3}\sqrt{z_0+1}  \\ &\times& \left [E\left(\frac{1-z_0}{1+z_0} \right) -z_0K\left(\frac{1-z_0}{1+z_0} \right)  \right]
+\mathcal O(\Lambda^{-1/2}) \,,   \nonumber 
\end{eqnarray}
The second integral we consider is the one in Eq. (\ref{intR}) with a cutoff in momenta 
\begin{eqnarray}
S(\Lambda)&=& 4\pi \int_m^{\Lambda} \frac{
 E  \sqrt{E ^2-m^2}   dE }
 {(2\pi)^4}  \\&\times&   \int _{C_F^{(f)}}    \frac{ k_{0}(1-g_2k_{0}) \;  dk_{0}}{g_2^4(k_0
-\omega_1 )^2 (k_0-\omega_2 )^2 (k_0-W_1 )^2 (k_0-W_2 ) ^2 }\,. \nonumber
\end{eqnarray}
We obtain in the limit $\Lambda \to \infty$:
\begin{eqnarray}
S(\Lambda)&=& -\frac{i \Lambda^{1/2}}{16 \sqrt{g}  \pi^2 }  \nonumber  \\  &+&\frac{i   }{16\pi^2g_2\sqrt{1+4g_2m}}
  \left[  (1+4g_2m) \right.  \nonumber \\ &\times& \left. 
   E\left(\frac{1-4g_2m}{1+4g_2m}\right) \right.  \nonumber   \\     & -& \left.4g_2m K\left(\frac{1-4g_2m}{1+4g_2m}\right)  \right] +\mathcal O(\Lambda^{-1/2})  
 \,.
\end{eqnarray}
We note that both integrals do not involve logarithmic
divergences and therefore we expect that dimensional regularization
gives finite results as well.


\end{document}